\documentclass[twocolumns]{pasj00}
\SetRunningHead{T. Izumi et al.}{ALMA observation of the active nucleus of NGC 1097}
\Received{2013/04/15}
\Accepted{2013/06/01}

\usepackage{times}

\begin{document}

\title{Submillimeter ALMA Observations of the Dense Gas in the Low-Luminosity Type-1 Active Nucleus of NGC 1097}
\author{Takuma \textsc{Izumi},\altaffilmark{1}
		Kotaro \textsc{Kohno},\altaffilmark{1,2}
		Sergio \textsc{Mart$\acute{\i}$n}, \altaffilmark{3}
		Daniel, \textsc{Espada}, \altaffilmark{4,5}
		Nanase \textsc{Harada}, \altaffilmark{6}
		Satoki \textsc{Matsushita}, \altaffilmark{7}
		Pei-Ying \textsc{Hsieh}, \altaffilmark{7,8}
		Jean L. \textsc{Turner}, \altaffilmark{9}
		David S. \textsc{Meier}, \altaffilmark{10,11}
		Eva \textsc{Schinnerer}, \altaffilmark{12}
		Masatoshi \textsc{Imanishi}, \altaffilmark{13}
		Yoichi \textsc{Tamura},\altaffilmark{1}
		Max T. \textsc{Curran}, \altaffilmark{14}
		Akihiro \textsc{Doi}, \altaffilmark{15}
		Kambiz \textsc{Fathi}, \altaffilmark{16,17}
		Melanie \textsc{Krips}, \altaffilmark{18}
		Andreas A. \textsc{Lundgren}, \altaffilmark{5}
		Naomasa \textsc{Nakai}, \altaffilmark{19}
		Taku \textsc{Nakajima}, \altaffilmark{20}
		Michael W. \textsc{Regan}, \altaffilmark{21}
		Kartik \textsc{Sheth}, \altaffilmark{22}
		Shuro \textsc{Takano}, \altaffilmark{14,23}
		Akio \textsc{Taniguchi}, \altaffilmark{1}
		Yuichi \textsc{Terashima}, \altaffilmark{24}
		Tomoka \textsc{Tosaki}, \altaffilmark{25}
		Tommy \textsc{Wiklind}, \altaffilmark{5}
		
\thanks{Last update: January 19, 2007}}
\altaffiltext{1}{Institute of Astronomy, The University of Tokyo, 2-21-1 Osawa, Mitaka, Tokyo 181-0015}
\email{takumaizumi@ioa.s.u-tokyo.ac.jp}
\altaffiltext{2}{Research Center for the Early Universe, The University of Tokyo, 7-3-1 Hongo, Bunkyo, Tokyo 113-0033}
\altaffiltext{3}{European Southern Observatory, Alonso de C\'{o}rdova 3107, Vitacura, Santiago, Chile}
\altaffiltext{4}{NAOJ Chile Observatory, National Astronomical Observatory of Japan, 2-21-1 Osawa, Mitaka, Tokyo 181-8588}
\altaffiltext{5}{Joint ALMA Observatory, Alonso de C\'{o}rdova 3107, Vitacura, Santiago, Chile}
\altaffiltext{6}{Max Planck Institute for Radio Astronomy, D-53121, Bonn, Germany}
\altaffiltext{7}{Academia Sinica, Institute of Astronomy \& Astrophysics, P.O. Box 23-141, Taipei 10617, Taiwan}
\altaffiltext{8}{Institute of Astronomy, National Central University, No. 300, Jhongda Road, Jhongli City, Taoyuan County 32001, Taiwan}
\altaffiltext{9}{Department of Physics and Astronomy, UCLA, 430 Portola Plaza, Los Angeles, Calfornia 90095-1547, USA}
\altaffiltext{10}{Department of Physics, New Mexico Institute of Mining and Technology, 801 Leroy Place, Soccoro, New Mexico 87801, USA}
\altaffiltext{11}{National Radio Astronomy Observatory, P.O. Box O, Soccoro, New Mexico 87801, USA}
\altaffiltext{12}{Max Planck Institute for Astronomy, K\"{o}nigstuhl 17, Heidelberg 69117, Germany}
\altaffiltext{13}{Subaru Telescope, National Astronomical Observatory of Japan, 650 North Afohoku Place, Hilo, Hawaii 96720, USA}
\altaffiltext{14}{Nobeyama Radio Observatory, NAOJ, Nobeyama, Minamimaki, Minamisaku, Nagano 384-1305}
\altaffiltext{15}{Institute of Space and Astronautical Science, 3-1-1 Yoshinodai, Chuo, Sagamihara 252-5210}
\altaffiltext{16}{Stockholm Observatory, Department of Astronomy, Stockholm University, AlbaNova Centre, 106 91 Stockholm, Sweden}
\altaffiltext{17}{Oskar Klein Centre for Cosmoparticle Physics, Stockholm University, 106 91 Stockholm, Sweden}
\altaffiltext{18}{Institute for Radio-Astronomy at Millimeter Wavelengths, Domaine Univ., 300 Rue de la Piscine, 38406 Saint Martin dfHeres, France}
\altaffiltext{19}{Institute of Physics, University of Tsukuba, 1-1-1 Ten-nodai, Tsukuba, Ibaraki 305-8571}
\altaffiltext{20}{Solar-Terrestrial Environment Laboratory, Nagoya University, Furo, Chikusa, Nagoya, Aichi 464-8601}
\altaffiltext{21}{Space Telescope Science Institute, 3700 San Martin Drive, Baltimore, Maryland 21218, USA}
\altaffiltext{22}{National Radio Astronomy Observatory, 520 Edgemont Road, Charlottesville, Virginia 22903, USA}
\altaffiltext{23}{Department of Astronomical Science, The Graduate University for Advanced Studies (Sokendai), Nobeyama, Minamimaki, Minamisaku, Nagano 384-1305, Japan}
\altaffiltext{24}{Department of Physics, Ehime University, Matsuyama, Ehime 790-8577}
\altaffiltext{25}{Department of Geoscience, Joetsu University of Education, Yamayashiki, Joetsu, Niigata 943-8512}


\maketitle

\begin{abstract}
We present the first 100 pc scale view of the dense molecular gas 
in the central $\sim$ 1.3 kpc of the type-1 Seyfert NGC 1097 traced by HCN ($J$ = 4--3) and HCO$^+$ ($J$ = 4--3) lines afforded with ALMA band 7. 
This galaxy shows significant HCN enhancement with respect to HCO$^+$ and CO in the low-$J$ transitions, 
which seems to be a common characteristic in AGN environments. 
Using the ALMA data, we study the characteristics of the dense gas around this AGN and search for the mechanism of HCN enhancement. 
We find a high HCN ($J$ = 4--3) to HCO$^+$ ($J$ = 4--3) line ratio in the nucleus. 
The upper limit of the brightness temperature ratio of HCN ($v_2$ = 1$^{1f}$, $J$ = 4--3) to HCN ($J$ = 4--3) is 0.08, 
which indicates that IR pumping does not significantly affect the pure rotational population in this nucleus. 
We also find a higher HCN ($J$ = 4--3) to CS ($J$ = 7--6) line ratio in NGC 1097 than in starburst galaxies, 
which is more than 12.7 on the brightness temperature scale. 
Combined from similar observations from other galaxies, we tentatively suggest that this ratio 
appears to be higher in AGN-host galaxies than in pure starburst ones similar to the widely used HCN to HCO$^+$ ratio. 
LTE and non-LTE modeling of the observed HCN and HCO$^+$ lines 
using $J$ = 4--3 and 1--0 data from ALMA, and $J$ = 3--2 data from SMA, 
reveals a high HCN to HCO$^+$ abundance ratio (5 $\leq$ [HCN]/[HCO$^+$] $\leq$ 20: non-LTE analysis) in the nucleus, 
and that the high-$J$ lines ($J$ = 4--3 and 3--2) are emitted from dense (10$^{4.5}$ cm$^{-3}$ $\leq$ $n_{\rm{H_2}}$ $\leq$ 10$^6$ cm$^{-3}$), 
hot (70 K $\leq$ $T_{\rm{kin}}$ $\leq$ 550 K) regions. 
Finally we propose that the $``${\it{high temperature chemistry}}$"$ is more plausible to explain the observed enhanced HCN emission in NGC 1097 than the pure gas phase PDR/XDR chemistry. 
 \end{abstract}

\section{Introduction}
The dense molecular medium plays various roles in the vicinity of active galactic nuclei (AGNs). 
The broad line region of an AGN is expected to be surrounded or obscured by a compact, 
dense dust torus on scales of $<$ 1 to a few tens of pc (e.g., Antonucci 1993). 
This dense torus could be a reservoir of fuel for nuclear activity, and could also be a site of massive star formation.
Thus, investigation of dense molecular gas is the key to the study of the nature 
of the underlying physical processes accompanying these activities (i.e., AGNs and starbursts). 
In addition, the feedback of activity onto the surrounding interstellar matter (ISM) may represent an important factor 
in the existence and evolution of the activity itself. 
Recent model calculations of the ISM predict that the various heating mechanisms will produce 
different signatures in molecular properties, e.g., photodissociation regions (PDRs) are formed by intense UV radiation from massive stars 
and X-ray dominated regions (XDRs) are formed near the very center of AGN 
(Maloney et al. 1996; Meijerink \& Spaans 2005;  Meijerink et al. 2007). 
Cosmic rays from supernovae (SNe) and the injection of mechanical energy 
induced by the AGN jet or SNe (mechanical heating) are also important to the chemical layouts 
(Matsushita et al. 2007; Loenen et al. 2008; Garc\'{\i}a-Burillo et al. 2010; Meijerink et al. 2011). 
Harada et al. (2010, 2013) modeled the chemical and thermal structure of the ISM around an AGN in terms of high temperature. 
To identify the different energy sources mm and submm spectroscopic observations are essential 
for probing buried AGNs in dusty nuclei and investigating their nature as these wavelengths do not suffer from dust extinction. 
Furthermore, nearby galaxies can serve as local templates of distant galaxies. 

With this in mind, many key molecules have been identified 
and proposed as useful diagnostic tools for the ISM in galaxies by mm and submm spectroscopic observations. 
For example, strong HCN ($J$ = 1--0) emission, which requires dense ($n_{\rm{H_2}} > 10^4$ cm$^{-3}$) 
environments for its collisional excitation, has been detected in the prototypical 
type-2 Seyfert NGC 1068 (Jackson et al. 1993; Tacconi et al. 1994; Helfer \& Blitz 1995; Kohno et al. 2008; Krips et al. 2012). 
Similar enhancements in low-luminosity Seyfert galaxies such as NGC 5194 (Kohno et al. 1996), 
NGC 1097 (Kohno et al. 2003), NGC 5033 (Kohno et al. 2005), and NGC 6951 (Kohno et al. 1999; Krips et al. 2007) 
have been reported. In these Seyfert nuclei, the HCN ($J$ = 1--0) to CO ($J$ = 1--0) integrated intensity ratios 
on the brightness temperature scale, $R_{\rm{HCN/CO}}$, are enhanced to approximately 0.4 -- 0.6, 
and the kinematics of the HCN line indicate that this dense molecular medium could be the outer envelope of 
the obscuring material (Jackson et al. 1993; Tacconi et al. 1994; Kohno et al. 1996; Krips et al. 2007; Davies, Mark, Sternberg 2012). 
While large $R_{\rm{HCN/CO}}$ ratios have been observed in AGNs (e.g., $\sim$ 0.5 in NGC 1068; Kohno et al. 2005), 
much smaller $R_{\rm{HCN/CO}}$ ($<$ 0.3) ratios are detected in pure starburst 
or composite (i.e., AGN + starburst) galaxies (e.g., M 82; Gao \& Solomon 2004a,b). 
Inactive galaxies have even lower ratios of $R_{\rm{HCN/CO}} < 0.1$. 
A similar trend has also been found in the ratio of HCN ($J$ = 1--0) to HCO$^+$ ($J$ = 1--0), $R_{\rm{HCN/HCO^+}}$ 
(e.g., Kohno et al. 1999, 2000, 2001, Kohno 2005). 
Therefore, $R_{\rm{HCN/CO}}$ vs $R_{\rm{HCN/HCO^+}}$ could be a useful discriminator between AGN and starburst activity, 
although there are some counter arguments that high $R_{\rm HCN/CO}$ ratios 
have also been observed in non-AGN galaxies (e.g., Snell et al. 2011; Costagliola et al. 2011) 
and low $R_{\rm HCN/HCO^+}$ ratios in AGN galaxies (e.g., Sani et al. 2012). 
However, the genuine presence of AGN in deeply obscured apparently non-AGN galaxies is debated even based on molecular observations 
(i.e., Arp 220: Gonz\'{a}lez-Alfonso et al. 2004; Rangwala et al. 2011; Mart\'{i}n et al. 2011). 

The cause of this HCN enhancement, on the other hand, is not clear because many different effects 
can contribute to an HCN enhancement in active environments, including higher gas opacities/densities and/or temperatures, 
non-standard molecular abundances caused by strong UV/X-ray radiation fields. 
In addition to these causes, a non-collisional excitation 
such as an IR pumping through the reradiation from UV/X-ray heated dust might be significant (e.g., Sakamoto et al. 2010; Aalto et al. 1995, 2002; Garci\'{a}-Carpio et al. 2006; Imanishi et al. 2009). 
Although the HCN molecule is usually used as a dense gas tracer because of its large dipole moment ($\mu$ = 3.0 Debye), 
it will not trace dense gas if another excitation mechanism exists that is faster 
than collisions with the H$_2$ molecule and independent of gas density. 
IR pumping is one such path, which is a transition among the molecular vibrationally excited states pumped by infrared radiation. 
The excitation can cause a transition between the vibrational excited and ground states, 
and the subsequent process can significantly increase the intensities of the rotational transitions (Carroll \& Goldsmith 1981). 

In the case of NGC 1068, several observations have demonstrated that the nuclear gas chemistry is dominated by 
X-ray radiation from the AGN yielding significantly different molecular abundances 
from starburst or quiescent environments (e.g., Usero et al. 2004; Garc$\acute{\i}$a-Burillo et al. 2010), 
and it has been proposed that the abundances of certain ions, radicals, and molecular species like HCN, 
can be enhanced in XDRs (Lepp \& Dalgarno 1996; Maloney et al. 1996; Meijerink \& Spaans 2005; Meijerink et al. 2007). 
However, it remains unclear whether these chemical models can fully explain the observations. 
Clearly, interferometric high angular resolution and high sensitivity observations are necessary 
to study the cause of HCN enhancement in AGN environments.

The CS molecule, on the other hand, is known to show little variation in abundance among galaxies with different activity (Mart\'{\i}n et al. 2009), 
which is confirmed in a variety of Galactic molecular clouds (Mart\'{\i}n et al. 2008). 
Therefore, this species would be a better dense gas reference to estimate 
the molecular abundance variations in galaxies. 

Our target galaxy, NGC 1097 is a nearby ($D = 14.5$ Mpc, Tully 1988) barred spiral galaxy classified as SB(s)b (de Vaucouleurs et al. 1991). 
It hosts a Seyfert 1 nucleus 
as evidenced by the double-peaked broad Balmer emission lines 
with time variability (FWHM $\sim$ 7500 km s$^{-1}$, Storchi-Bergmann et al. 1997; Schimoia et al. 2012). 
This type-1 AGN is surrounded by a circumnuclear starburst ring with a radius of 10$''$ 
or 700 pc (Barth et al. 1995; Quillen et al. 1995), which is prominent at various wavelengths, 
including radio (Hummel et al. 1987), mid-infrared (Kotilainen et al. 2000; Reunanen et al. 2010; Sheth et al. 2010;  Beir\~{a}o et al. 2012; Kondo et al. 2012), 
and soft X-rays (P\'erez-Olea \& Colina 1996). 
Although some papers argue that the nuclear star clusters are the dominant contributor to the heating of the dust in the nucleus in mid-IR (e.g., Mason et al. 2007; Asmus et al. 2011), 
the detection of a hard X-ray source at the nucleus (Nemmen et al. 2006) also supports
the presence of a genuine active nucleus. 
The estimated luminosity of the AGN in NGC 1097 is a rather low 
(low--luminosity AGN = LLAGN; $L_{\rm{2-10keV}} = 4.4 \times 10^{40}$ erg s$^{-1}$, 
$L_{\rm{bol}}$ = 8.6 $\times$ 10$^{41}$ erg s$^{-1}$ at $D = 14.5$ Mpc). 
A timing analysis of X-ray light curves implies that a super-massive black hole 
does exist in NGC 1097 (Awaki et al. 2001). 
Two pairs of huge ($\sim$ a few 10 kpc scale) optical jets have also been reported, 
yet their nature is unclear (e.g., Wehrle et al. 1997; Galianni et al. 2010). 
All relevant properties are summarized in Table \ref{1097property}. 
The molecular condensation at the nucleus described above shows an elevated $R_{\rm{HCN/HCO^+}}$ ratio 
of $\sim$ 2 for both the $J$ = 1--0 and $J$ = 3--2 transitions (Kohno et al. 2003; Hsieh et al. 2012). 
The CO ($J$ = 2--1) to CO ($J$ = 1--0) line ratio reaches 1.8 (Hsieh et al. 2008), 
which also indicates the existence of the strong heating source in the nucleus. 
These observed anomalous molecular line ratios suggest that 
the physical and chemical properties of the molecular phase are likely dominated by the AGN in the center of NGC 1097. 
The star formation rate (SFR) in the circumnuclear starburst ring is high 
(5 $M_{\odot}$ yr$^{-1}$ from the extinction corrected H$\alpha$ luminosity -- Hummel et al. 1987). 
The ring is connected to the nucleus by dusty spiral features (Prieto et al. 2005).
These structures make NGC 1097 an ideal laboratory to study AGN vs. starburst effects 
on molecular material at high resolution, as well as the connection between the dense material 
and the central engine. 

In this paper, we present the first 100 pc scale view of the type-1 Seyfert nucleus in NGC 1097 traced by submillimeter dense gas tracers such as 
HCN ($J$ = 4--3), HCO$^+$ ($J$ = 4--3), CS ($J$ = 7--6), CO ($J$ = 3--2), and HCN ($v_2 = 1^{1f}$, $J$ = 4--3) 
afforded by the {\it{Atacama Large Millimeter/submillimeter Array (ALMA)}}, though Hsieh et al. (2011) present comparable resolution observations of CO ($J$ = 2--1) line from this galaxy. 
The achieved angular resolution of $\sim 1''.5$ or $\sim 100$ pc at a distance of 14.5 Mpc matches the distance up to which heating 
due to the AGN is expected to be effective (Schleicher et al. 2010), and it is sufficient to separate the emission from 
the AGN and the surrounding circumnuclear starburst ring in NGC 1097. 
This angular resolution is a factor $\sim$ 7 better in area than previous HCN and HCO$^+$ observations by Hsieh et al. (2012). 
Compared to low-$J$ millimeter lines, these high-$J$ submillimeter lines can trace denser and/or hotter region with higher angular resolution in general, 
which means these high-$J$ lines are more suitable to investigate the properties of AGNs. 
In addition, these high-$J$ lines can be easily observed in high redshift objects as they are redshifted to the millimeter range. 

We describe in Section 2 the specifications of our observations. 
An 860 $\mu$m continuum map and astrometric information are shown in Section 3. 
We display channel maps of significantly detected lines in Section 4 and a full band 7 spectrum is shown in Section 5. 
We briefly discuss the dense molecular gas kinematics in the nucleus based on a simple assumption in Section 6. 
Some interesting molecular line ratios are discussed in Section 7. 
The physical condition of the dense molecular gas is investigated both under LTE and non-LTE conditions in Section 8. 
Possible interpretations of the observed properties are discussed in Section 9, 
and our main conclusions are summarized in Section 10. 

\begin{table*}[h]
\begin{minipage}{\textwidth}
  \caption{Properties of NGC 1097}\label{1097property}
  \begin{center}
    \begin{tabular}{ccc}
    \hline\hline
    Parameter & Value & Reference  \\ \hline
    Morphology & SB(s)b & 1 \\
    Nuclear activity & Type 1 Seyfert & 2 \\
    Position of nucleus: &  & 3 \\
    $\alpha$(J2000.0) & 02$^\mathrm{h}$46$^\mathrm{m}$18$^\mathrm{s}$.96 &  \\
    $\delta$(J2000.0) & -30$^\circ$16$'$28$''$.9&  \\
    $V_{\rm sys}$ (km s$^{-1}$) & 1271 & 4 \\
    $D_{25} \times d_{25}$(arcmin) & 9.3 $\times$ 6.3 & 1 \\
    Position angle (deg) & 130 & 1 \\
    Inclination angle (deg) & 46 & 5 \\
    Adopted distance (Mpc) & 14.5 & 6 \\
    Linear scale (pc arcsec$^{-1}$) & 70 & 6 \\
    $L_{\rm{2-10keV}}$ (erg s$^{-1}$) & $4.4 \times 10^{40}$ & 7 \\
    SFR(ring) ($\rm{M_{\odot}}$ yr$^{-1}$) & 5 & 3 \\ \hline
    \end{tabular}
  \end{center}
  \footnotetext{{\bf{References.}} --- (1) de Vaucouleurs et al. 1991; (2) Storchi-Bergmann et al. 1993; (3) Hummel et al. 1987;\\
   (4)Koribalski et al. 2004; (5) Ondrechen et al. 1989; (6) Tully 1988; (7) Nemmen et al. 2006}
  \end{minipage}
\end{table*}

\section{Observation and Data Reduction}
We observed NGC 1097 with the band 7 receiver on ALMA 
using the 2SB dual-polarization setup, as a cycle 0 early science program (ID = 2011.0.00108.S; PI = K. Kohno) 
on November 5$^{\rm th}$ and 6$^{\rm th}$ 2011. 
The observations were conducted in a single pointing with 18$''$ field of view, 
centered at the intensity-weighted centroid of 
the combined {\it{2MASS}} ($J$ + $H$ + $K_s$) image ($J+H+K_s$ peak, hereafter; Jarrett et al. 2003), 
R.A. (J2000.0) = $02^{\rm{h}} 46^{\rm{m}} 19^{\rm{s}}.05$, Dec. (J2000.0) = $-30^{\circ} 16' 29''.7$. 
The receiver was tuned to cover the redshifted HCN ($J$ = 4--3), HCO$^+$ ($J$ = 4--3) and HCN ($v_2=1^{1f}$, $J$ = 4--3) 
lines in the upper sideband (USB), and H${}^{13}$CN ($J$ = 4--3), HC${}^{15}$N ($J$ = 4--3) and CS ($J$ = 7--6) with the lower sideband (LSB). 
Each spectral window has a bandwidth of 1.875 GHz, and two spectral windows were set to 
each sideband to achieve a total frequency coverage of $\sim$ 7.5 GHz in this observation. 
The spectral resolution is 0.488 MHz per channel, but 20 channels were binned together to improve the S/N ratio, 
which results in a final spectral resolution of 9.8 MHz ($\sim$ 8.5 km s$^{-1}$). 
The assumed heliocentric systemic velocity was 1271 km s$^{-1}$, 
which was used to calculate sky frequencies (based on HI observation; Koribalski et al. 2004). 
In the following part of this paper, we express velocities in the optical convention. 
The difference to the radio convention is $\sim$ 5.4 km s$^{-1}$ for this galaxy. 
The observations were performed with 14 antennas in the first run and 15 antennas in the second, both in the compact configuration. 
The resulting $uv$ range covers $\sim$ 14 -- 134 k$\lambda$. 
Weather conditions were good throughout the observation with system temperatures of 150 -- 200 K. 
The bandpass and phase were calibrated with J0522-364 and J0334-401, respectively. 
This phase calibrator is 13.9 deg away from NGC 1097. 
Callisto was also observed as a flux calibrator. 
Parameters of the observation and other specifications are summarized in Table \ref{specifications}. 

The calibration of the data was done with CASA (McMullin et al. 2007; Petry \& the CASA Development Team 2012) in standard manners. 
The image was reconstructed with the CASA task \verb|CLEAN| (the number of iteration was 500, and no mask was used) 
and analyzed with MIRIAD (Sault et al. 1995). 
Using natural weighting, the achieved synthesized beams are 1$''$.50 $\times$ 1$''$.20, P.A. = -72.4$^\circ$ for HCN ($J$ = 4--3), 
1$''$.49 $\times$ 1$''$.18, P.A. = -71.3$^\circ$ for HCO$^+$ ($J$ = 4--3), for example. 
We achieved rms noises in channel maps of $\sim$ 2.1 mJy beam$^{-1}$ and $\sim$ 2.3 mJy beam$^{-1}$ in the LSB and the USB, 
at a velocity resolution of 8.6 km s$^{-1}$ and 8.3 km s$^{-1}$, respectively. 
An image of the continuum emission was obtained using the CASA task \verb|CLEAN| by averaging channels free of line emission. 
The rms noise for the continuum image is 0.37 mJy beam$^{-1}$ centered at 348.6 GHz (860 $\mu$m) after combining both the LSB and USB  data, 
and the synthesized beam is 1$''$.52 $\times$ 1$''$.21, P.A. = 108$^\circ$. 
This continuum emission was subtracted in the $uv$-plane before making line maps. 
Throughout this paper, the pixel scale of all maps is set to 0$''$.3/pix, 
and displayed errors indicate only statistical ones unless mentioned otherwise. 
When we include the systematic error of the absolute flux calibration, 
$\sim$ 10 \% accuracy is estimated for both sidebands. 

\begin{table*}
  \caption{ALMA Band 7 Observation Parameters}\label{specifications}
  \begin{center}
    \begin{tabular}{ccc}
    \hline\hline
       Parameters &  &  \\ \hline
       Date & \multicolumn{2}{c}{2011 Nov 5, 6} \\
       Configuration & \multicolumn{2}{c}{compact} \\
       Phase center: & & \\
       R.A. & \multicolumn{2}{c}{$\alpha$(J2000.0) = $\rm{02^h 46^m 19^s.05}$} \\
       Dec. & \multicolumn{2}{c}{$\delta$(J2000.0) = $\rm{-30^{\circ} 16' 29''.7}$} \\
       Primary beams & \multicolumn{2}{c}{18$''$}  \\
       No. of antennas & \multicolumn{2}{c}{14 (first day), 15 (second day)} \\
       Projected baseline range [k$\lambda$] & \multicolumn{2}{c}{14 --134} \\
       Bandpass calibrator & \multicolumn{2}{c}{J0522-364} \\
       Absolute flux calibrator & \multicolumn{2}{c}{Callisto} \\ 
       Gain calibrator & \multicolumn{2}{c}{J0334-401} \\ 
        $T_{\rm{sys}}$ [K] & \multicolumn{2}{c}{150 -- 200}\\
        & & \\
        & LSB & USB \\ \hline
       Frequency coverage [GHz] & 340.98 -- 344.43 & 352.61 -- 356.18 \\
       Velocity resolution [km s$^{-1}$] & 8.6 & 8.3 \\
       Central frequency of each spectral window [GHz] & 342.07, 343.66 & 353.67, 355.41 \\
       Rms noise in channel maps [mJy beam$^{-1}$] & 2.1 & 2.3 \\
       Conversion factor [K (Jy beam$^{-1}$)$^{-1}$] & 5.43 & 5.40 \\ \hline
    \end{tabular}
    \end{center}
\end{table*}

\section{Continuum map}
\begin{figure*}
  \begin{center}
    \FigureFile(150mm,150mm){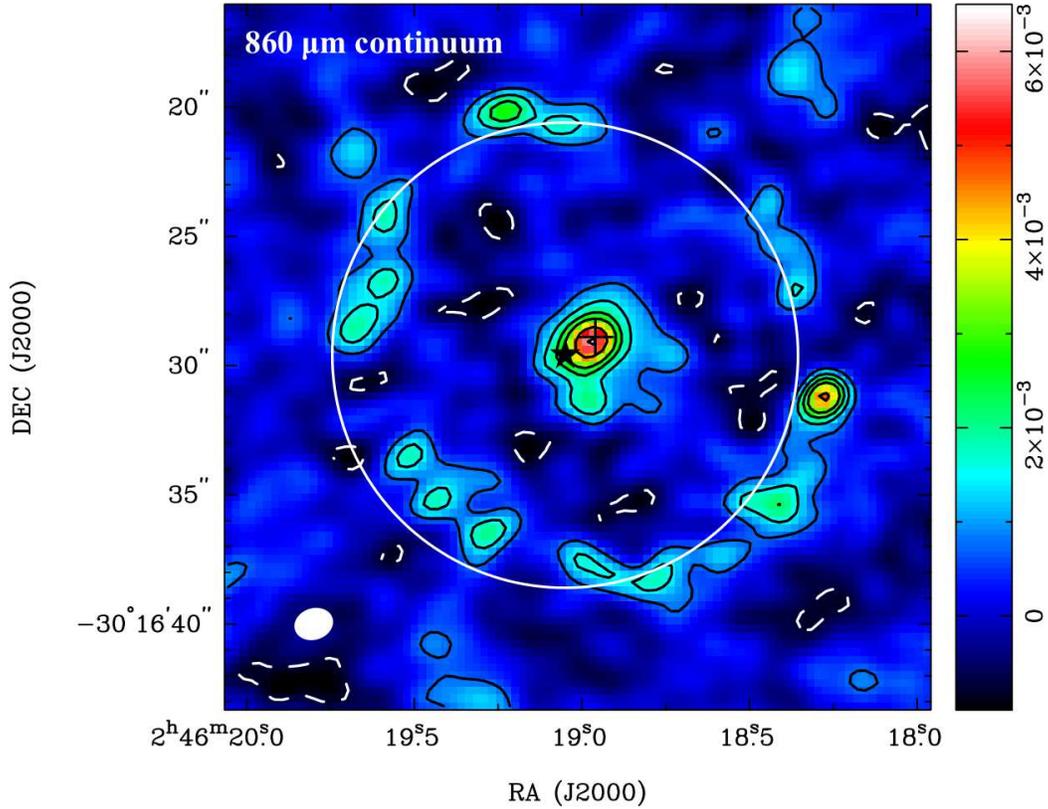}
  \end{center}
  \caption{Continuum map towards the nucleus of NGC 1097 at 860 $\mu$m. 
  The units of the scale bar is in Jy beam$^{-1}$. 
  Contours are -2, 2, 4, 6, 8, 12, and 16 $\sigma$ (black-solid contours are positive and dashed-white contours are negative ones, respectively), 
  where 1 $\sigma$ = 0.37 mJy beam$^{-1}$ or 2.0 mK in brightness temperature scale. 
  The maximum is 6.13 mJy beam$^{-1}$ or 33 mK, located at the nucleus. 
  The white-filled ellipse indicates the beam ellipse at 860 $\mu$m (1$''$.52 $\times$ 1$''$.21, P.A.=108$^{\circ}$). 
  The cross and the star indicate the peak positions of VLA 6 cm, and intensity-weighted centroid of the combined 2MASS ($J$ + $H$ + $K_s$) image, respectively. 
  It is obvious that the 6 cm peak position and that of 860 $\mu$m are identical with each other. 
  The field of view of ALMA at this frequency is 18$''$ (indicated by the large white circle), and the attenuation 
  due to the primary beam pattern of each element antenna is not corrected in this map, 
  i.e., no primary beam correction has been applied. }
  \label{B7cont}
\end{figure*}

Figure \ref{B7cont} shows the 860 $\mu$m continuum map towards the central 30$''$ (2.1 kpc) region of NGC 1097. 
Continuum emission is detected from both the nucleus and the circumnuclear starburst ring. 
The continuum peak (6.13 $\pm$ 0.37 mJy) is 
at R.A. = $\rm{02^h 46^m 18^s.97}$, Dec. = $\rm{-30^{\circ} 16' 29''.2}$, 
which precisely coincides with the 6 cm continuum peak, however, 
it is shifted $\sim$ 1$''$.5 from the $J+H+K_s$ peak (see also Table \ref{peaks}). 
The precise match between 6 cm and 860 $\mu$m continuum peaks supports the accuracy of the astrometry of ALMA, 
despite the fact that the phase calibrator used (J0334-401) is 13.9 deg away from NGC 1097, which can be causing a significant 
position error if the array has a significant baseline error. 

The peak position of the 2MASS $J+H+K_s$ combined image (Jarrett et al. 2003) in Figure 1 
is significantly shifted from the AGN position (i.e., 860 $\mu$m and 6 cm peak position), even when taking errors into account. 
On the other hand, the peak positions of individual $J$, $H$, and $K_s$ band image extracted from 2MASS Image Service\footnote{http://irsa.ipac.caltech.edu/applications/2MASS/IM/} 
precisely coincide with that of the AGN. 
The reason of this disagreement between 2MASS combined and individual images is not clear, and to reveal it is out of the scope of this paper. 
However, if we regard that the astrometry of the individual images is accurate, 
previously revealed compact NIR source (Prieto et al. 2005) 
and compact nuclear star formation traced by its hot dust continuum (Davies et al. 2007) in the nuclear region appear to be co-located with the AGN, 
although its star formation activity is estimated to be very low (SFR $\sim$ 0.1 $M_\odot$ yr$^{-1}$ within the central 22 pc radius; calculated from Table 2 in Davies et al. 2007). 

We fit a 2-dimensional Gaussian to the nuclear 860 $\mu$m continuum source 
using only $>$ 5 $\sigma$ emission to derive the intrinsic size of the nucleus using the MIRIAD task \verb|IMFIT|. 
The derived source size is 2$''$.01 $\times$ 1$''$.35 (141 pc $\times$ 95 pc) with P.A. = -26$^{\circ}$.1, thus it is resolved. 
A spectral energy distribution (SED) of the nucleus including radio to submillimeter continuum indicates that the 860 $\mu$m continuum 
from the nucleus is consistent with thermal emission from dust (Matsushita et al. in prep.). 

The overall structure of the continuum emitting region in the circumnuclear starburst ring 
consists of several clumps, which are also visible in maps of CO ($J$ = 2--1), (3--2) observed with the SMA (Hsieh et al. 2008, 2011). 
These structures appear very reminiscent of those visible in the 18.7 $\mu$m map in Roenanen et al. (2010), 
although the true morphology and the extent of the 860 $\mu$m continuum is not clear due to the primary beam attenuation. 
Taking into these things account, we consider that the 860 $\mu$m continuum emission from the circumnuclear starburst ring 
would be tracing the massive star-formation. 

\begin{table*}
  \caption{Peak positions of continuum and HCN line emissions}\label{peaks}
  \begin{center}
    \begin{tabular}{cccc}
    \hline\hline
      Wavelength & R.A. (J2000.0) & Dec. (J2000.0) & Ref. \\ \hline
      6 cm & 02$^{\rm h}$46$^{\rm m}$18$^{\rm s}$.96 ($\pm$ 0$^{\rm s}$.02) & -30$^{\circ}$16$'$28$''$.9 ($\pm$ 0$''$.2) & Hummel et al. 1987 \\
      $J+H+K_s$ & 02$^{\rm h}$46$^{\rm m}$19$^{\rm s}$.05 ($\pm$0$^{\rm s}$.02) & -30$^{\circ}$16$'$29$''$.7 ($\pm$0$''$.3) & Jarrett et al. 2003 \\
      860 $\mu$m & 02$^{\rm h}$46$^{\rm m}$18$^{\rm s}$.97 ($\pm$ 0$^{\rm s}$.03) & -30$^{\circ}$16$'$29$''$.2 ($\pm$ 0$''$.6) & This work \\
      HCN ($J$ =4--3) & 02$^{\rm h}$46$^{\rm m}$18$^{\rm s}$.98 ($\pm$ 0$^{\rm s}$.03) & -30$^{\circ}$16$'$29$''$.0 ($\pm$ 0$''$.6) & This work \\ \hline
          \end{tabular}
    \end{center}
\end{table*}

\section{Channel maps and integrated intensities}
\begin{figure*}
  \begin{center}
    \FigureFile(180mm,180mm){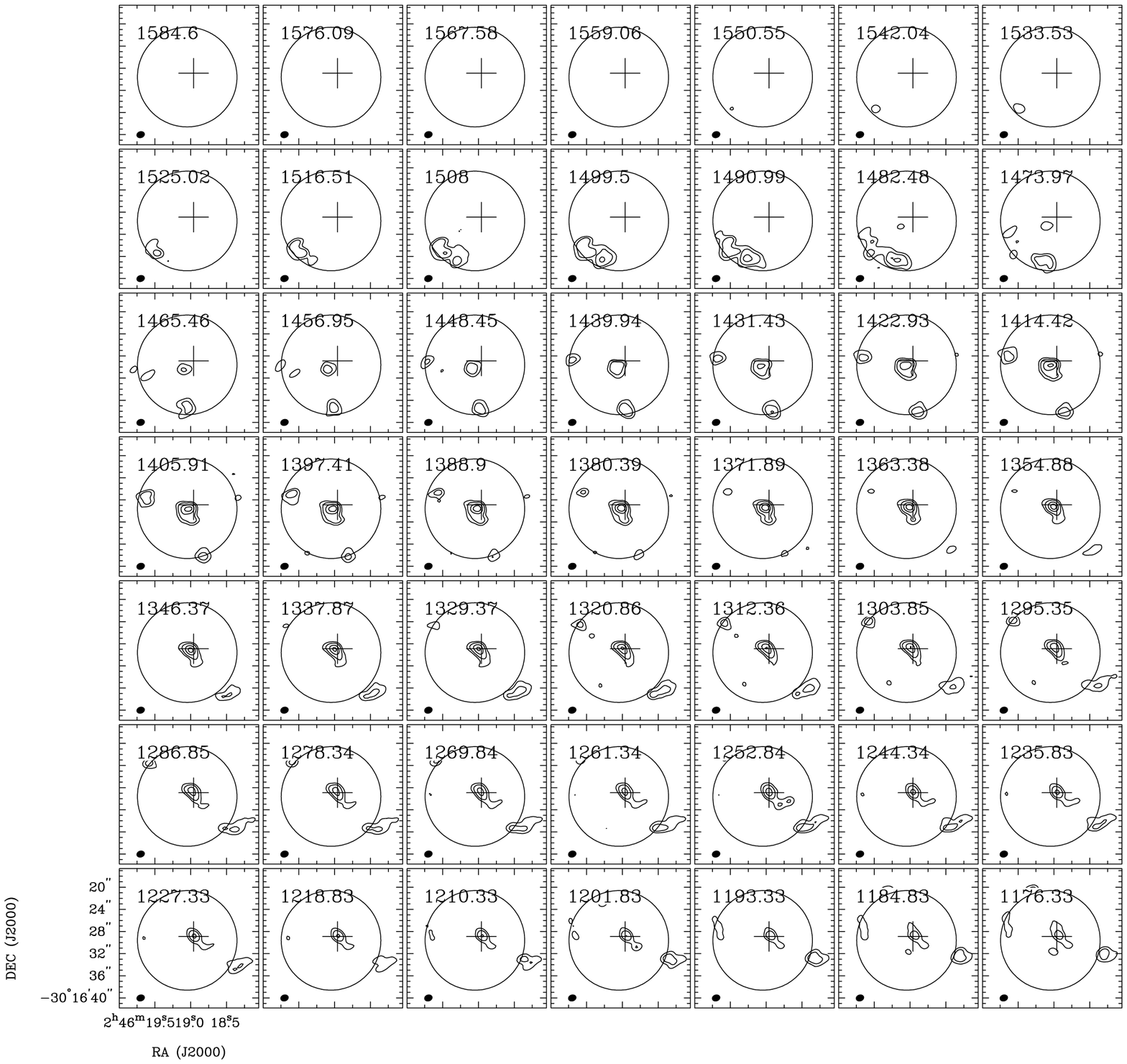}
  \end{center}
  \caption{
  The velocity channel maps of the CO ($J$ = 3--2) line emission in the central 25$'' \times$ 25$''$ 
  (1.75 kpc $\times$ 1.75 kpc at $D = 14.5$ Mpc) region of NGC 1097. 
  The central cross in each channel indicates the position of the nucleus defined by the peak position of the 860 $\mu$m continuum, 
  which corresponds to that of the 6 cm peak. 
  The velocity width of each channel is 8.5 km s$^{-1}$, and the central velocities ($V_{\rm LSR}$ in km s$^{-1}$) of each channel map are indicated. 
  The beam size is 1$''$.55 $\times$ 1$''$.22 with P.A. = 109$^{\circ}$, plotted in the bottom-left corner of each channel. 
  The field of view of ALMA at this frequency (18$''$) is indicated by the large black circle. 
  Contour levels are 50, 100, 150, 200, 250, and 300 $\sigma$, where 1 $\sigma$ = 2.51 mJy beam$^{-1}$ or 13.6 mK in $T_b$. 
  Only the positive contours with high levels are plotted to see the overall structure. 
  Note that we could not observe the entire line emission due to our correlator setup. 
  Attenuation due to the primary beam pattern is not corrected. 
  }
  \label{COchmap}
\end{figure*}

\begin{figure*}
  \begin{center}
    \FigureFile(180mm,180mm){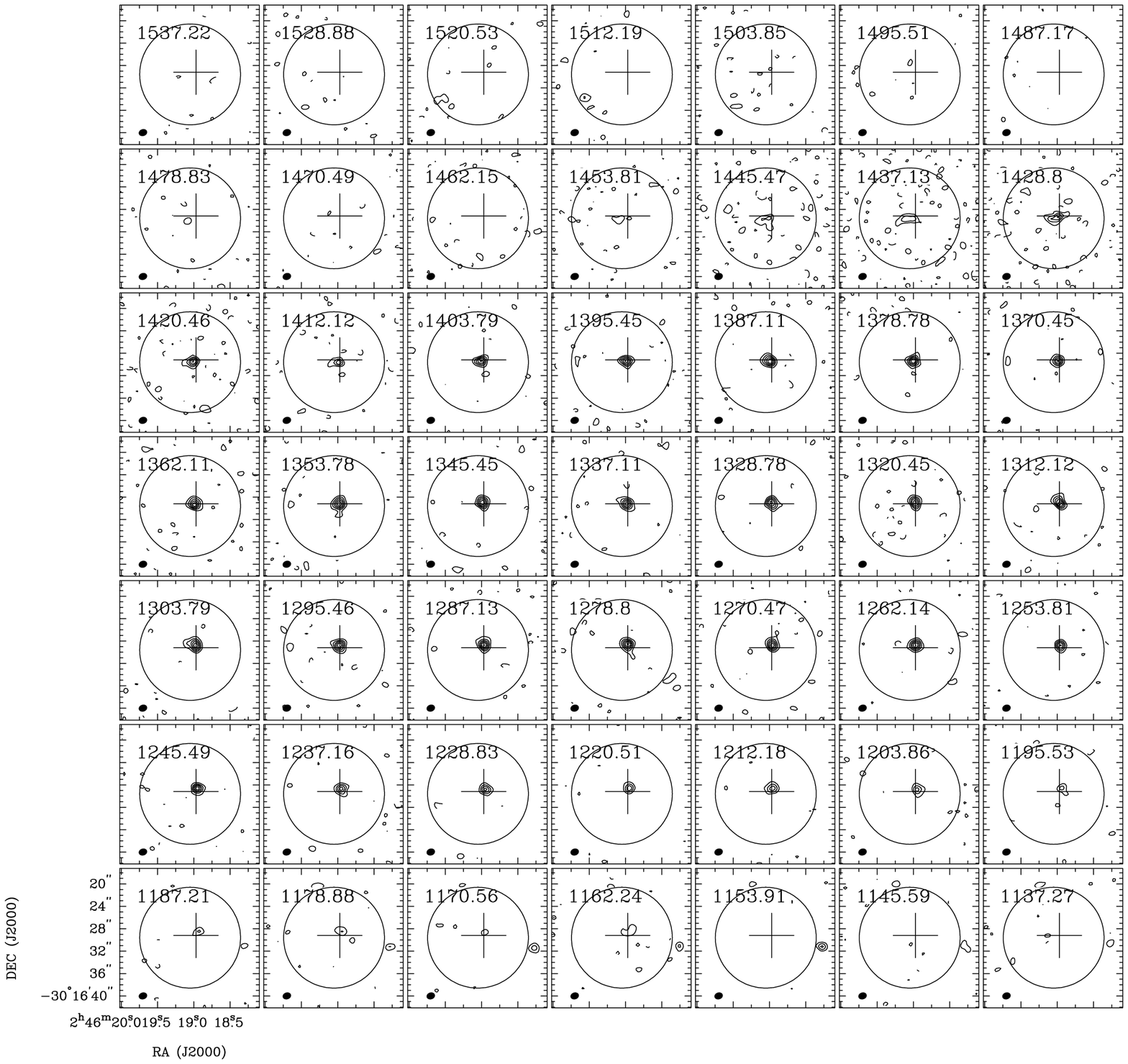}
  \end{center}
  \caption{
  The velocity channel maps of the HCN ($J$ = 4--3) emission in the central 25$''$ $\times$ 25$''$ (1.75 kpc $\times$ 1.75 kpc at $D$ = 14.5 Mpc). 
  The velocity width of each channel is 8.3 km s$^{-1}$, 
  and the central velocities ($V_{\rm LSR}$ in km s$^{-1}$) are labeled in each channel map. 
  The 860 $\mu$m continuum peak is also indicated by a cross. 
  The beam size is 1$''$.50 $\times$ 1$''$.20, P.A. = -72.4$^\circ$, plotted in the bottom-left corner of each channel. 
  The field of view of ALMA at this frequency (18$''$) is indicated by the large black circle. 
  Contour levels are -3, 3, 6, 9, 12, 15, and 18 $\sigma$,where 1 $\sigma$ = 2.27 mJy beam$^{-1}$ or 12.3 mK in $T_b$. 
  The negative contours are plotted with the dashed lines. 
  Attenuation due to the primary beam pattern is not corrected.
  }
  \label{HCNchmap}
\end{figure*}

\begin{figure*}
  \begin{center}
    \FigureFile(180mm,180mm){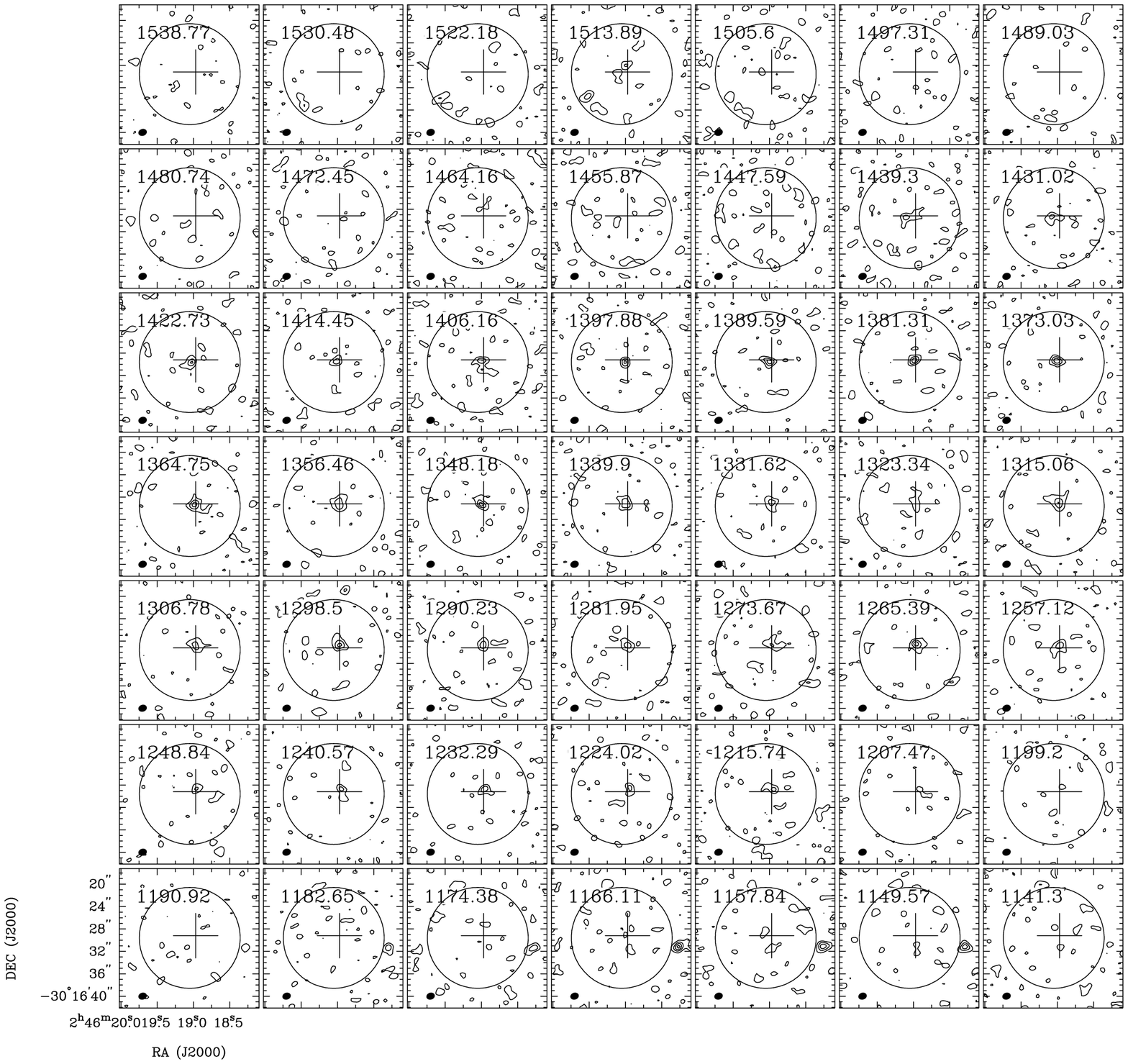}
  \end{center}
  \caption{
  The velocity channel maps of HCO$^+$ ($J$ = 4--3) emission in the central 25$''$ $\times$ 25$''$ (1.75 kpc $\times$ 1.75 kpc at $D$ = 14.5 Mpc). 
  The velocity width of each channel is 8.3 km s$^{-1}$, 
  and the central velocities ($V_{\rm LSR}$ in km s$^{-1}$) are labeled. 
  The 860 $\mu$m continuum peak is also indicated by a cross.
  The beam size is 1$''$.49 $\times$ 1$''$.18, P.A. = -71.3$^\circ$, plotted in the bottom-left corner of each channel. 
  The field of view of ALMA at this frequency (18$''$) is indicated by the large black circle. 
  Contour levels are -3, 3, 6, 9, and 12 $\sigma$,where 1 $\sigma$ = 2.39 mJy beam$^{-1}$ or 12.9 mK in $T_b$. 
  The negative contours are plotted with the dashed lines. 
  Attenuation due to the primary beam pattern is not corrected.
  }
  \label{HCOPchmap}
\end{figure*}

Figures \ref{COchmap} to \ref{HCOPchmap} show the channel maps of the CO ($J$ = 3--2), HCN ($J$ = 4--3), and HCO$^+$ ($J$ = 4--3) line emission 
in the central 25$'' \times$ 25$''$ region (1.75 kpc $\times$ 1.75 kpc) of NGC 1097. 
We could not cover the whole CO ($J$ = 3--2) line due to spectral setting restrictions for ALMA cycle 0 observations, but the detected emission 
in each channel is extremely significant (peak flux density is $\sim$ 300 $\sigma$). 
HCN ($J$ = 4--3) emission was detected ($>$ 3 $\sigma$) over a velocity range of $V_{\rm{LSR}}$ = 1160--1450 km s$^{-1}$, 
and HCO$^+$ ($J$ = 4--3) emission was detected ($>$ 3 $\sigma$) over $V_{\rm{LSR}}$ = 1215--1440 km s$^{-1}$ in the nucleus. 
Note this is the first detections of HCN ($J$ = 4--3) and HCO$^+$ ($J$ = 4--3) emission from NGC 1097. 
The velocity ranges at a level of 3 $\sigma$ are roughly consistent among the $J$ = 4--3, 3--2 and 1--0 transitions (Kohno et al. 2003; Hsieh et al. 2012). 
The differing velocity ranges between the HCN and HCO$^+$ molecules are probably due to the lower S/N ratio of HCO$^+$. 
It should be noted that these velocity ranges are 200 -- 300 km s$^{-1}$ smaller 
than that of CO ($J$ = 1--0), (2--1) and (3--2) (Kohno et al. 2003; Hsieh et al. 2008, 2011), 
but previous CO observations had larger beam sizes, thus they might have also observed an outer diffuse, high-velocity component. 

In the channel maps, it is clear that the CO ($J$ = 3--2) emission originates 
from locations of the Seyfert nucleus and the circumnuclear starburst ring, 
whereas strong HCN ($J$ = 4--3) and HCO$^+$ ($J$ = 4--3) emissions originate primarily from the nucleus ($r \leq 100$ pc). 
Note however, the primary beam size is 18$''$ in this observation, which is comparable to that of the circumnuclear starburst ring, 
thus the primary beam attenuation is severe in the region. 
Details of the circumnuclear starburst ring will be discussed in future papers. 
Therefore in this paper, we focus our attention on the nuclear region, especially at the peak position of the 860 $\mu$m continuum 
because our interests are the physical and chemical properties of dense molecular gas in the nucleus. 

\begin{figure*}
  \begin{center}
    \FigureFile(180mm,180mm){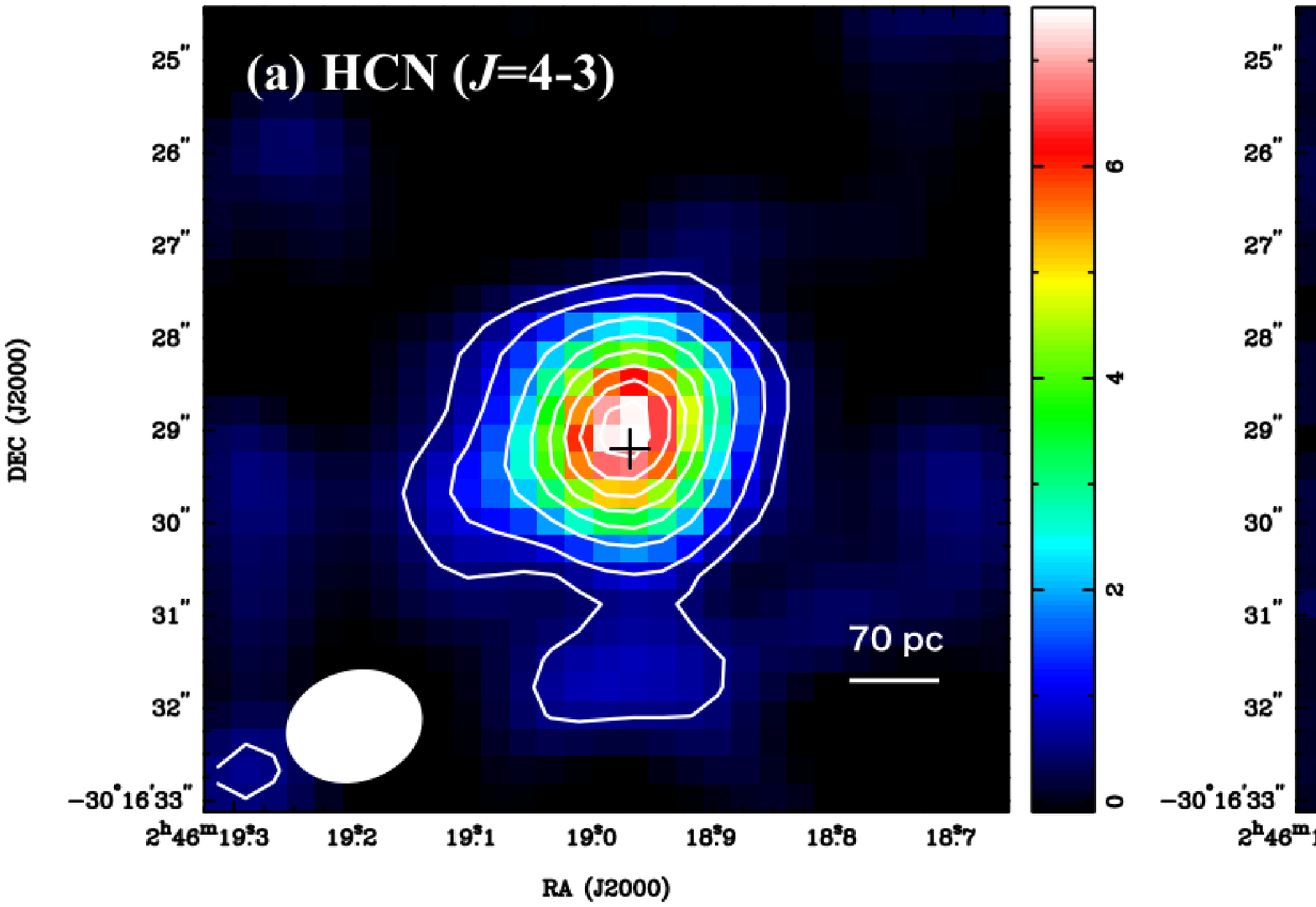}
  \end{center}
  \caption{Integrated intensity maps in the central 8$''$ $\times$ 8$''$ (560 pc $\times$ 560 pc) region of NGC 1097, 
  derived by calculating the zeroth moment of the ALMA data cubes. 
  The cross indicate the peak position of the 860 $\mu$m continuum. 
  Both maps are shown on the same intensity scale to allow for easy comparison. 
  (a) Integrated intensity map of the HCN ($J$ = 4--3) emission over a velocity range 
  from $V_{\rm{LSR}}$ = 1160 to 1450 km s$^{-1}$. 
  The value at the 860 $\mu$m peak position is 7.5 Jy beam$^{-1}$ km s$^{-1}$ 
  and the source size is estimated to be 1$''$.34 $\times$ 1$''$.04 with P.A. = -9.1$^\circ$. 
  The beam size is 1$''$.50 $\times$ 1$''$.20 with P.A. = -72.4$^\circ$. 
  The contour levels are 5, 10, 20, ${}_{\cdots}$, 70 $\sigma$, where 1 $\sigma$ is 0.11 Jy beam$^{-1}$ km s$^{-1}$ or 0.60 K km s$^{-1}$. 
  (b) Integrated intensity map of the HCO$^+$ ($J$ = 4--3) emission over a velocity range 
  from $V_{\rm{LSR}}$ = 1215 to 1440 km s$^{-1}$. 
  The value at the 860 $\mu$m peak position is 3.7 Jy beam$^{-1}$ km s$^{-1}$ 
  and the source size is estimated to be 1$''$.31 $\times$ 1$''$.06 with P.A. = -14.7$^\circ$. 
   The beam size is 1$''$.49 $\times$ 1$''$.18 with P.A. = -71.3$^\circ$. 
   The contour levels are 5, 10, 15, ${}_{\cdots}$, 30 $\sigma$, where 1 $\sigma$ is 0.11 Jy beam$^{-1}$ km s$^{-1}$ or 0.60 K km s$^{-1}$. 
  }
  \label{mom0}
\end{figure*}

The HCN ($J$ = 4--3) and HCO$^+$ ($J$ = 4--3) velocity integrated intensity maps are displayed in Figure \ref{mom0}. 
These images are made by calculating the zeroth moment from three dimensional data cubes, using the MIRIAD task \verb|MOMENT|. 
To minimize the contribution from noise, we computed these moment maps by using channels which contain $> $ 3 $\sigma$ emission only, 
where 1 $\sigma$ = 2.27 mJy beam$^{-1}$ for HCN ($J$ = 4--3) and 2.39 mJy beam$^{-1}$ for HCO$^+$ ($J$ = 4--3), respectively. 
The corresponding velocity ranges are already listed above. 
The integrated intensity of HCN ($J$ = 4--3) at the 860 $\mu$m peak position is 7.5 $\pm$ 0.1 Jy beam$^{-1}$ km s$^{-1}$, 
and the intrinsic source size is estimated to be 1$''$.34 $\times$ 1$''$.04 with P.A. = -9.1$^\circ$ 
(note the beam size is 1$''$.50 $\times$ 1$''$.20). 
In the case of HCO$^+$ ($J$ = 4--3), the integrated intensity is 3.7 $\pm$ 0.1 Jy beam$^{-1}$ km s$^{-1}$, 
and the estimated intrinsic source size is 1$''$.31 $\times$ 1$''$.06 with P.A. = -14.7$^\circ$. 

\section{Band 7 spectra}
\begin{figure*}
  \begin{center}
    \FigureFile(170mm,170mm){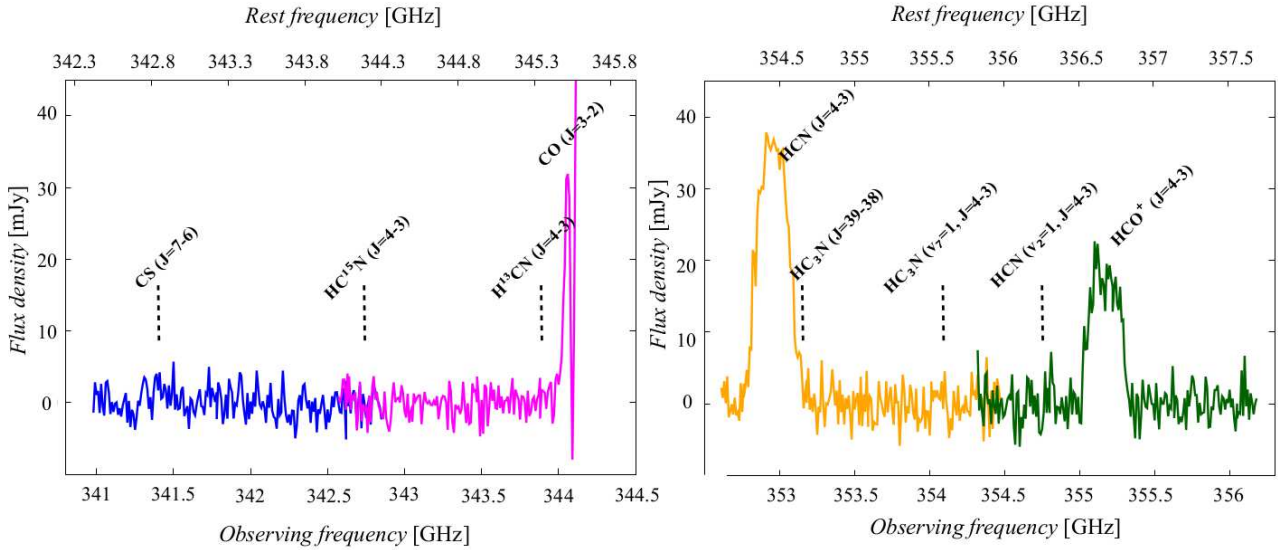}
  \end{center}
  \caption{ALMA Band 7 spectra at the 860 $\mu$m peak of NGC 1097. 
  Indicated spectral lines are marked at the systemic velocity of the galaxy (1271 km s$^{-1}$; Koribalski et al. 2004). 
  We detected CO ($J$ = 3--2), HCN ($J$ = 4--3) and HCO$^+$ ($J$ = 4--3) with $>$ 3 $\sigma$ significance. 
  Other lines are undetected by this observation. 
  Full line names and line parameters are listed given in Table \ref{line}. 
  The velocity resolutions and rms noise levels are in Table \ref{specifications}. 
  Colors indicate the different spectral windows (spw 0, 1, 2, and 3, from left to right) 
  and the continuum emission has already been subtracted.}
  \label{B7spec}
\end{figure*}

The spectra and emission parameters extracted at the 860 $\mu$m peak position are presented in Figure \ref{B7spec} and Table \ref{line}. 
Note that we only claim that a line is real if it shows up as a $>$ 3 $\sigma$ feature. 
The strongest line is CO ($J$ = 3--2) followed by HCN ($J$ = 4--3) and HCO$^+$ ($J$ = 4--3). 
Some CO ($J$ = 3--2) emission lies outside the observed spectral window as already explained, 
thus we exclude the parameters of the CO ($J$ = 3--2) line from Table \ref{line}. 
Other lines including CS ($J$ = 7--6), HC${}^{15}$N ($J$ = 4--3), H${}^{13}$CN ($J$ = 4--3), 
HC${}_3$N ($J$ = 39--38) and HCN ($v_2$ = 1$^{1f}$, $J$ = 4--3) are undetected (i.e., $<$ 3 $\sigma$). 
Note that vibrationally excited HCN is in its bending mode ($v_2$ = 1), and one part of the $l$-doublets ($l$ = 1$f$). 
The counterpart ($l$ = 1$e$) is very close to the ground state HCN line, therefore, it is indistinguishable due to the large line widths in this galaxy. 
Also, HC${}_3$N ($J$ = 39--38) is so close to HCN ($J$ = 4--3) that they are blended. 

\begin{figure*}
  \begin{center}
    \FigureFile(160mm,160mm){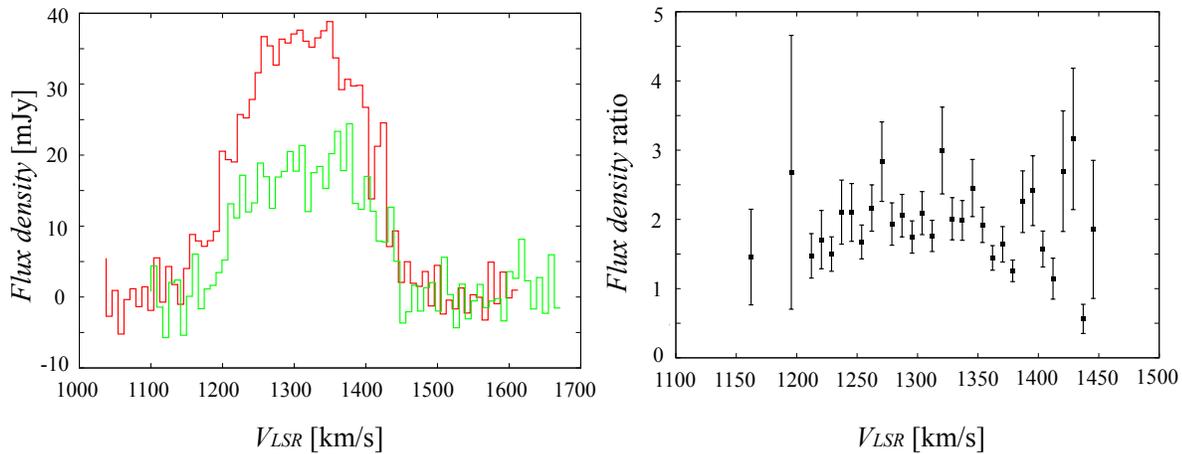}
  \end{center}
  \caption{
  ({\it{left}}) HCN ($J$ = 4--3) ($red$) and HCO$^+$ ($J$ = 4--3) ($green$) spectra 
  at the 860 $\mu$m peak position are displayed as functions of $V_{\rm{LSR}}$. 
  Emissions above 3 $\sigma$ has been detected in the velocity range for $V_{\rm{LSR}}$ = 1160 -- 1450 km s$^{-1}$ 
  for HCN ($J$ = 4--3), and 1215 --1440 km $^{-1}$ in HCO$^+$ ($J$ = 4--3). 
  We calculated the integrated intensities in Table \ref{line} over these velocity ranges.  
  ({\it{right}}) HCN ($J$ = 4--3) to HCO$^+$ ($J$ = 4--3) flux density ratio as a function of $V_{\rm{LSR}}$. 
  We plot only channels for which a robust determination of the ratio can be made. 
  The ratio is almost constant within the range where both lines are detected significantly. 
  Only the statistical error is considered here. 
  		}
  \label{flux}
\end{figure*}

To derive the integrated intensities of HCN ($J$ = 4--3) and HCO$^+$ ($J$ = 4--3) in Table \ref{line}, 
we used the zeroth moment maps in the previous subsection, 
not results of Gaussian fitting to the spectra, because as shown in Figures \ref{B7spec} and \ref{flux}, 
the spectra of these lines show slightly asymmetric profiles and it is difficult to fit a Gaussian profile to them to achieve firm peak flux densities and line widths. 
This spectral asymmetry (blue shifted part is fainter than the redshifted part in both lines) 
in the nucleus of NGC 1097 has also been claimed in lower transitions of these molecules and CO (Kohno et al. 2003; Hsieh et al. 2012), 
and it is likely due to multiple underlying components. 
Higher angular resolution observations are required to further investigate this feature. 
Figure \ref{flux} also shows the HCN ($J$ = 4--3) to HCO$^+$ ($J$ = 4--3) flux density ratio as a function of $V_{\rm{LSR}}$. 
We found that the ratio is almost constant within the range where the both lines are detected significantly. 
This result could be understandable, for example, if multiple small clouds with similar physical conditions within the nucleus were observed. 
The velocity widths ($\Delta$ $v$) of these lines are defined as the channel width (8.3 km s$^{-1}$) 
times  the number of channels above the half maximum (this corresponds to a FWHM in the case of a Gaussian profile).
We assume that the velocity error is $\pm$ 1 channel (8.3 km s$^{-1}$). 
The total integrated flux within the central $r$ $\leq$ 2$''$.5 (175 pc) is 13.6 $\pm$ 0.2 Jy km s$^{-1}$ for HCN ($J$ = 4--3), 
and 7.8 $\pm$ 0.2 Jy km s$^{-1}$ for HCO$^+$ ($J$ = 4--3), respectively. 
Even at a degraded $\sim$ 50 km s$^{-1}$ resolution, 
no emission from the lines indicated in Figure \ref{B7spec} was detected 
other than HCN ($J$ = 4--3), HCO$^+$ ($J$ = 4--3) and CO ($J$ = 3--2). 
Upper limits to the integrated intensities were derived for these undetected transitions 
assuming a Gaussian profile with FWHM similar to the $\Delta$$v$ derived for HCN ($J$ = 4--3). 
By using these derived values, for example, the H${}^{12}$CN to H${}^{13}$CN line ratio is $>$ 12.7 (3 $\sigma$) on the brightness temperature scale, 
indicating that our main target line HCN ($J$ = 4--3) has $\tau$ $<$ a few and is not severely optically thick 
considering the $^{12}$C /$^{13}$C isotopic ratio obtained so far 
(e.g., $\sim$ 50 in Galactic sources; Lucas \& Liszt 1998, $>$ 40 in starburst galaxies; Mart\'{\i}n et al. 2010).

It should be noted that among the lines listed in Table \ref{line}, HCN ($J$ = 4--3) and HCO$^+$ ($J$ = 4--3) 
could be observed in high redshift objects by using ALMA as these lines are relatively strong (detectable) 
and they are still within the frequency coverage of ALMA after being redshifted (if $z$ is less than $\sim$ 3.2). 
Therefore, we also list the line luminosities of HCN ($J$ = 4--3) and HCO$^+$ ($J$ = 4--3) in the central 100 pc region in Table \ref{luminosity}, 
which can be compared with high redshift objects. 
We used equations (2) and (3) in Solomon \& Vanden Bout (2005) for this calculation. 

\begin{table*}
 \caption{Emission properties of NGC 1097}\label{line}
  \begin{minipage}{\textwidth}
  \begin{center}
    \begin{tabular}{ccccccc}
    \hline\hline
       Emission & $\nu_{\rm{rest}}$ & $E_u/k_{\rm B}$ & Peak flux & $\Delta$$v$ & $I$ & $I$ \\
        & [GHz] & [K] & [mJy beam$^{-1}$] & [km s$^{-1}$] & [Jy beam$^{-1}$ km s$^{-1}$] & [K km s$^{-1}$] \\
                (1) & (2) & (3) & (4) & (5) & (6) & (7)  \\ \hline
       HCO$^+$ ($J$ = 4--3) & 356.734 & 42.8 & 24.4$\pm$2.4 & 174$\pm$8.3 & 3.7$\pm$0.1 & 20.2$\pm$0.5  \\
       HCN ($v_2=1^{1f}$, $J$ = 4--3) & 356.256 & 1067.1 & $<$ 2.8 & - & $<$ 0.60 & $<$ 3.2   \\
       HC${}_3$N ($v_7=1^{1e}$, $J$ = 39--38) & 355.566 & 662.2 & $<$ 2.8 & - & $<$ 0.60 & $<$ 3.2   \\
       HC${}_3$N ($J$ = 39--38) & 354.697 & 340.5 & $<$ 2.8 & - & $<$ 0.60 & $<$ 3.2  \\
       HCN ($J$ = 4--3) & 354.505 & 42.5 & 38.8$\pm$2.3 & 199$\pm$8.3 & 7.5$\pm$0.1 & 40.5$\pm$0.5 \\
       H{}$^{13}$CN ($J$ = 4--3) & 345.340 & 41.4 & $<$ 2.6 & - & $<$ 0.55 & $<$ 3.2 \\
       HC{}$^{15}$N ($J$ = 4--3) & 344.200 & 41.3 & $<$ 2.6 & - & $<$ 0.55 & $<$ 3.2 \\
       CS ($J$ = 7--6) & 342.883 & 65.8 & $<$ 2.6 & - & $<$ 0.55 & $<$ 3.2 \\ 
       860 $\mu$m continuum & $\cdots$ & $\cdots$ & 6.13$\pm$0.37 & $\cdots$ & $\cdots$ & $\cdots$ \\ \hline
    \end{tabular}
    \end{center}
    \footnotetext{{\bf{Notes.}} Column 1: Full line name. Column 2: Rest frequency. Column 3: Upper energy level. 
    Column 4: peak flux density of line and continuum emission at the 860 $\mu$m continuum peak in mJy beam$^{-1}$. 
    For HCN ($J$ = 4--3) and HCO$^+$ ($J$ = 4--3), errors were estimated from the adjacent emission--free channels. 
    For other lines, 6 channels were binned to degrade the velocity resolution into $\sim$ 50 km s$^{-1}$ to improve the S/N ratio, 
    and  3 $\sigma$ upper limits estimated from adjacent emission--free channels are shown.  
    For continuum, an error was estimated from emission--free areas in the continuum channel. 
    Column 5: Velocity width defined as the channel width (8.3 km s$^{-1}$) times the number of channels above the half maximum of each line 
    (note that this is not a FWHM). 
    Column 6: Velocity integrated intensity in Jy beam$^{-1}$ km s$^{-1}$. 
    For the non-detected lines, we used a 3 $\sigma$ upper limit flux density and $\Delta$v of HCN ($J$ = 4--3) as a FWHM for calculation. 
    Column 7: Velocity integrated intensity in K km s$^{-1}$; 
    In this Table, only the statistical error is shown. 
    The systematic error of absolute flux calibration is estimated to be $\sim$ 10 \%.
    }
    \end{minipage}
\end{table*}

\begin{table*}[h]
  \caption{HCN($J$ = 4--3) and HCO$^+$ ($J$ = 4--3) line luminosities.}
  \label{luminosity}
  \begin{minipage}{\textwidth}
  \begin{center}
    \begin{tabular}{ccccc}
    \hline\hline
        & $L_{\rm{HCN}}$ [$L_{\odot}$] & $L_{\rm{HCO^+}}$ [$L_{\odot}$] & $L'_{\rm{HCN}}$ [K km s$^{-1}$ pc$^2$] & $L'_{\rm{HCO^+}}$ [K km s$^{-1}$ pc$^2$] \\
         & (1) & (2) & (3) & (4)\\ \hline
        860 $\mu$m peak & 577$\pm$8 & 291$\pm$8 & (4.05$\pm$0.05)$\times$10$^5$ & (2.00$\pm$0.05)$\times$10$^5$ \\ 
        $r$ $<$ 175 pc & 1049$\pm$16 & 608$\pm$16 & (7.37$\pm$0.10)$\times$10$^5$ & (4.17$\pm$0.10)$\times$10$^5$ \\ \hline
    \end{tabular}
    \end{center}
    \footnotetext
    {
    {{\bf{Notes.}} Column 1, 2: Line luminosity in [$L_{\odot}$] unit. Column 3, 4: Line luminosity in [K km s$^{-1}$ pc$^2$] unit. 
    Row 1: Line luminosity within the synthesized beam centered at the 860 $\mu$m continuum peak. Row 2: Line luminosity within the central $r$ $<$ 175 pc region. 
    }}
    \end{minipage}
\end{table*}

\section{Dense molecular gas kinematics}
\begin{figure*}
  \begin{center}
    \FigureFile(160mm,160mm){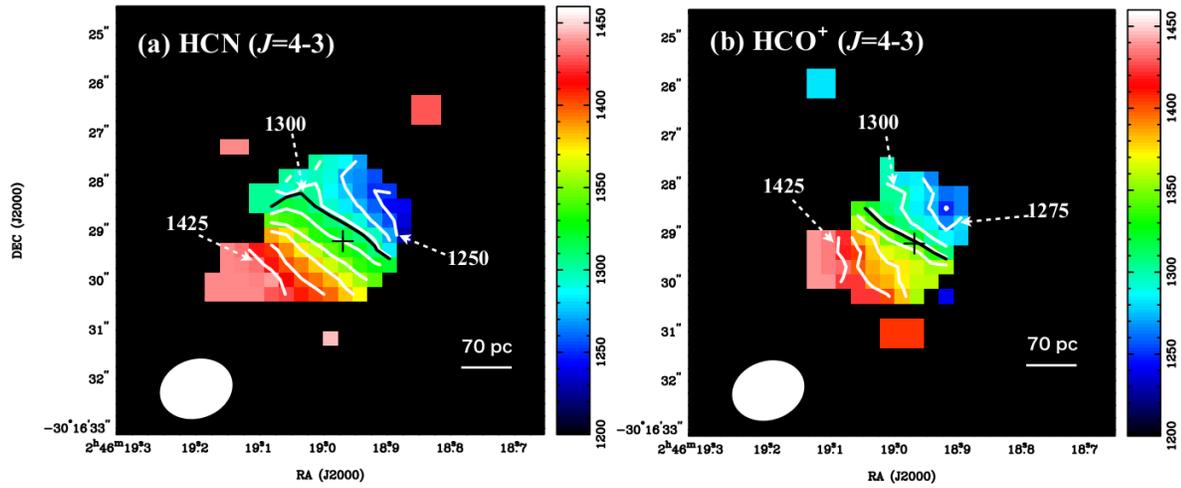}
  \end{center}
  \caption{
  Intensity--weighted mean velocity maps of HCN ($J$ = 4--3) and HCO$^+$ ($J$ = 4--3) 
  in the central 8$''$ $\times$ 8$''$ (560 pc $\times$ 560 pc) region of NGC 1097. 
  The cross indicates the peak positions of 860 $\mu$m continuum. 
  No significant deviation from circular rotation is visible at this angular resolution. 
  (a) Intensity--weighted mean velocity map of HCN ($J$ = 4--3) emission. 
  The contour interval is 25 km s$^{-1}$, and contours are labeled by the value of $V_{\rm{LSR}}$. 
  The median velocity, 1303 km s$^{-1}$, is also displayed as a thick black contour. 
  The beam size is 1$''$.50 $\times$ 1$''$.20 with P.A. = -72.4$^\circ$. 
  (b) Intensity-weighted mean velocity map of HCO$^+$ ($J$ = 4--3) emission. 
  The contour interval is 25 km s$^{-1}$, and contours are labeled by the value of $V_{\rm{LSR}}$. 
  The median velocity, 1327 km s$^{-1}$, is also displayed as a thick black contour. 
  The beam size is 1$''$.49 $\times$ 1$''$.18 with P.A. = -71.3$^\circ$. 
  }
  \label{mom1}
\end{figure*}

\begin{figure*}
  \begin{center}
    \FigureFile(90mm,90mm){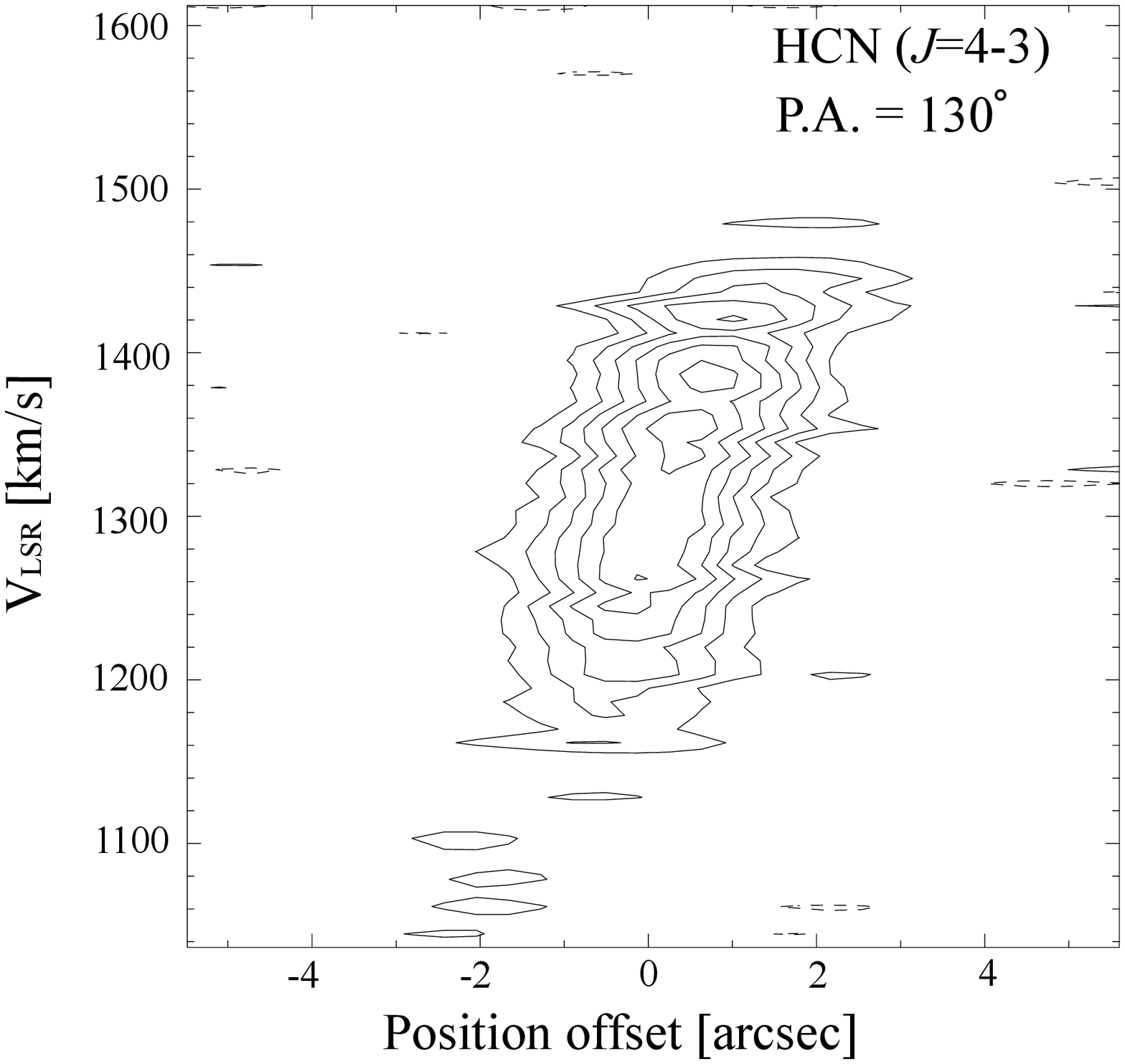}
  \end{center}
  \caption{Position--velocity diagram (PV diagram) of HCN ($J$ = 4--3) emission in the nucleus of NGC 1097 along the kinematic major axis (P.A. = 130$^\circ$). 
  The center is the 860 $\mu$m peak position. 
  Contour levels are 15, 30, 45, 60, 75 and 90\% of the maximum (38.8 mJy beam$^{-1}$), thus the 15\% contour corresponds to 2.6 $\sigma$. 
  The HCN emission is concentrated towards the nucleus. 
    }
  \label{pv}
\end{figure*}

\begin{figure*}
  \begin{center}
    \FigureFile(160mm,160mm){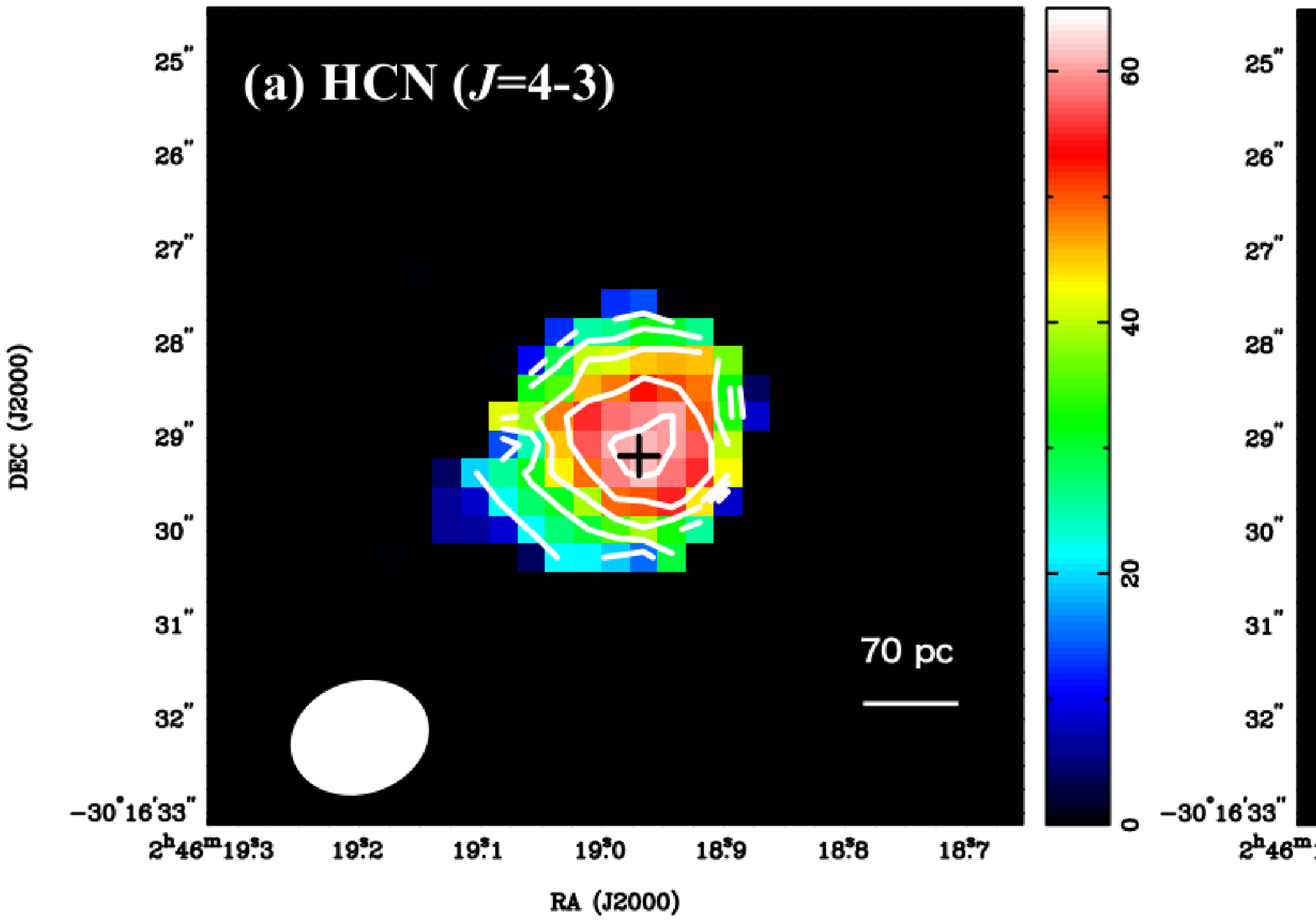}
  \end{center}
  \caption{
  Intensity--weighted velocity dispersion maps of HCN ($J$ = 4--3) and HCO$^+$ ($J$ = 4--3) 
  in the central 8$''$ $\times$ 8$''$ (560 pc $\times$ 560 pc) region of NGC 1097. 
  The cross indicates the peak position of 860 $\mu$m continuum. 
  (a) Intensity--weighted velocity dispersion map of HCN ($J$ = 4--3) emission. 
  The contour interval is 10 km s$^{-1}$, starting from 20 km s$^{-1}$,
  and the value at the 860 $\mu$m peak position is 61.1 km s$^{-1}$. 
  The beam size is 1$''$.50 $\times$ 1$''$.20 with P.A. = -72.4$^\circ$. 
  (b) Intensity-weighted velocity dispersion map of HCO$^+$ ($J$ = 4--3) emission. 
  The contour interval is 10 km s$^{-1}$, starting from 20 km s$^{-1}$, 
  and the value at the 860 $\mu$m peak position is 58.4 km s$^{-1}$. 
  The beam size is 1$''$.49 $\times$ 1$''$.18 with P.A. = -71.3$^\circ$. 
  }
  \label{mom2}
\end{figure*}

Figure \ref{mom1} shows intensity--weighted mean velocity maps of HCN ($J$ = 4--3) and HCO$^+$ ($J$ = 4--3) emissions. 
These maps were made by computing the first moment of each data cube with 5 $\sigma$ clipping using the MIRIAD task \verb|MOMENT|. 
Overall structures in both maps roughly exhibit a pure circular rotation in the central region of NGC 1097, 
and no significant deviation from it is readily apparent at this angular resolution. 
The median velocity of the HCN ($J$ = 4--3) map is 1303 km s$^{-1}$, and that of the HCO$^+$ ($J$ = 4--3) map is 1327 km s$^{-1}$. 
Clearly, the 860 $\mu$m continuum peak, and therefore the 6 cm continuum peak, match well with the median velocity, 
which implies that the Seyfert 1 nucleus is located at the dynamical center. 
Both of these values are consistent with those derived from a Fourier decomposition analysis of this HCN ($J$ = 4--3) data by Fathi et al. (2013). 
A position-velocity diagram (PV diagram) of the HCN ($J$ = 4--3) line emission along the major axis (P.A. = 130$^\circ$) of NGC 1097 is shown in Figure \ref{pv}. 
It is confirmed that the strong HCN emission is concentrated towards the nucleus and rotation is clearly visible (the velocity gradient is $\sim$ 2.3 km s$^{-1}$ pc$^{-1}$). 
Figure \ref{mom2} shows the intensity--weighted velocity dispersion maps of 
the HCN ($J$ = 4--3) and HCO$^+$ ($J$ = 4--3) emission along the line of sight, 
derived by calculating the second moment of each data cube with 5 $\sigma$ clipping. 
Note that the second moment produces a map of standard deviation of a Gaussian, not a FWHM by MIRIAD. 
It should also be noted that these maps contain both the intrinsic velocity dispersion 
of the gas and the gradient of rotation velocity within the observing beam. 
Therefore, we assume throughout the paper that the significantly increased velocity dispersion seen near the 860 $\mu$m continuum peak 
is due to the steeply rising in the rotation curve in the central region of the galaxy. 
 
Assuming a thin disk with Keplerian rotation, 
the dynamical mass within a radius $r$ is calculated as 
\begin{eqnarray}
\nonumber M_{\rm{dyn}} &=& \frac{rv^2(r)}{G}\cdot (\sin i)^{-2} \\
&=& 2.3 \times 10^2 \Biggl(\frac{r}{\rm{pc}}\Biggr) \Biggl[\frac{v(r)}{\rm{km\hspace{1mm} s^{-1}}} \Biggr]^2 (\sin i)^{-2} M_{\rm{\odot}}, 
\end{eqnarray}
where $v(r)$ the rotation velocity at radius $r$ from the center 
and $i$ is the inclination angle to correct for velocity projection. 
Using the estimated upper limit on the source size (1$''$.34 $\times$ 1$''$.04 = 94 pc $\times$ 73 pc), we set the radius to $r$ $\sim$ 40 pc, 
and we adopt an inclination angle of 35$^\circ$ (Fathi et al. 2006). 
This inclination angle is similar to the value derived by Storchi-Bergmann et al. (2003) for the nuclear accretion disk, 
but $\sim$ 10$^\circ$ smaller than the value derived from the large scale gas kinematics by Ondrechen et al. (1989).
By using these values, the enclosed dynamical mass within 40 pc is estimated to be $M_{\rm{dyn}}$ $\sim$ $2.8 \times 10^8 M_{\rm{\odot}}$. 
This value is within a factor of a few of previous measurements of the dynamical mass of 1.4 $\times$ 10$^8$ $M_{\rm \odot}$ based on higher angular resolution stellar kinematics (Davies et al. 2007), 
 and that of 9.5 $\times$ 10$^7$ $M_{\rm \odot}$ based on molecular gas kinematics (Hicks et al. 2009) in this galaxy. 
In addition, our value (and previously derived ones) is also within a factor of a few of a nuclear black hole mass of $\sim (1.2\pm0.2) \times 10^8 M_{\rm {\odot}}$, which is 
estimated via the central stellar velocity dispersion of the bulge using the CaII near-IR triplet lines (Lewis \& Eracleous 2006), 
although our estimation is based on a velocity measurement well outside the sphere of influence of 
the supermassive black hole in NGC 1097 (the radius is $\sim$ 13 pc, Lewis \& Eracleous 2006; Peebles 1972). 

Note that the estimation we present in this section is based on 
the simple assumptions of Keplerian rotation. 
In addition, the dynamical mass estimated here is just an lower limit because the rotation velocity may be underestimated due to the beam smearing effect. 
Therefore, the dynamical mass would change if we adopt different conditions and/or different values. 
For example, Fathi et al. (2013) assume a thin disk structure and isolate the rotational velocities contributing 
to the observed velocity field, and derive an enclosed dynamical mass within 40 pc radius of 8.0 $\times$ 10$^6$ $M_\odot$. 

\section{Molecular line ratios in the nucleus of NGC 1097}
Here, we first discuss whether IR-pumping affects the rotational population of HCN emission lines in NGC 1097. 
Next, we propose a tentative diagnostic diagram to investigate whether AGN contribute or not to the observed line emission. 

\subsection{Vibrationally excited HCN}
The first vibrationally excited state of HCN is its bending state ($v_2$ = 1), 1024 K above the ground. 
Considering the temperature for any known infrared source, we can expect that 
only transitions through this state will contribute significantly to the total HCN rotational population. 
Historically, the HCN ($v_2=1^{1f}$, $J$ = 4--3) emission has been detected in several Galactic sources, such as Ori-KL (Schilke et al. 1997). 
However, in the case of extragalactic sources, despite the increasing number of HCN observations, 
the convincing detection of vibrationally excited HCN is limited so far to 
the luminous infrared galaxy, NGC 4418 (Sakamoto et al. 2010), which has a infrared luminous 
and dust-shrouded nucleus ($L_{\rm{8-1000\mu m}}$=1.3$\times$10$^{11}L_{\odot}$; Dudley \& Wynn-Williams 1997). 
This is because extragalactic sources that can emit vibrationally excited HCN are intrinsically IR luminous ones, such as (U)LIRGs, 
(indeed, a vibrationally excited emission of HC$_3$N has also been observed in NGC 4418 
(Sakamoto et al. 2010; Costagliola et al. 2010) and Arp 220 (Mart\'{\i}n et al. 2011)), 
however, molecular line widths in such sources are so large that the HCN ($v_2=1^{1f}$, $J$ = 4--3) emission is severely blended with the adjacent HCO$^+$ ($J$ = 4--3) emission. 
Therefore, it is still controversial whether such radiative pumping is indeed significant in extragalactic HCN observations 
as physical scale we probe is much larger than for Galactic sources. 
Thus it is much more difficult to achieve sufficient IR flux in over the typical several hundreds pc observed in extragalactic sources. 
In the case of NGC 4418, the HCN ($v_2=1^{1f}$, $J$ = 4--3) to HCN ($J$ = 4--3) integrated intensity ratio 
on the brightness temperature scale ($\equiv$ $R_{\rm{HCNv/HCN}}$, hereafter) reaches 0.22 and Sakamoto et al. (2010) 
concluded that the high vibrational temperature derived from the detection is the consequence of 
IR-pumping, HCN does not simply trace dense gas in the nucleus of that galaxy. 
The detected vibrationally excited emission in NGC 4418 has been claimed to be most likely tracing the warmest gas 
in extremely  compact star forming regions within heavily obscured environments, 
though a hot component associated to a buried AGN can not be excluded. 

Here we present in Table \ref{HCNv_table} the $R_{\rm{HCNv/HCN}}$ of NGC 1097, 
accompanied by NGC 4418, Ori-KL and R-CrA for comparison. 
The $R_{\rm{HCNv/HCN}}$ in NGC 1097 is at least three times smaller than that in NGC 4418, 
and seems to be comparable to the ratio in Ori-KL. 
Note however, although the integrated intensity ratio is 0.052 in Ori-KL (Schilke et al. 1997), the peak intensity ratio increases to 0.14, which is much higher than NGC1097. 
One possible explanation for this discrepancy is that the HCN ($v_2=1^{1f}$, $J$ = 4--3) and HCN ($J$ = 4--3) lines 
are mainly emitted from different regions within the observing beam in Ori-KL, which results in the different line widths and integrated intensities. 
Therefore, we suggest that IR-pumping is not effective in Ori-KL at a 0.05 pc scale, 
but it will be more significant at smaller scales where molecular clouds and spot regions emitting the vibrationally excited HCN can be resolved. 
Considering those described above and the low $L_{\rm IR}$ (8.6 $\times$ 10$^7$ $L_{\odot}$, Prieto et al. 2010), 
which is $\sim$ 0.1 \% of that of NGC 4418, we expected that IR-pumping 
is not effective in the nucleus of NGC 1097 at a 100 pc scale, 
and that this mechanism does not affect the HCN rotational transitions. 
A similar conclusion was obtained from observation of submillimeter HCN lines in nearby starburst galaxies (Jackson et al. 1995). 
Therefore, we treat the HCN ($J$ = 4--3) emission as the consequence of the purely rotational transition, hereafter. 

\begin{table*}[h]
  \caption{HCN($v_2=1^{1f}$, $J$ = 4--3) / HCN($J$ = 4--3) integrated intensity ratio in brightness temperature scale.}\label{HCNv_table}
  \begin{minipage}{\textwidth}
  \begin{center}
    \begin{tabular}{ccccc}
    \hline\hline
        & NGC 1097 & NGC 4418 & Ori-KL & R-CrA \\ \hline
        Type & LLAGN & LIRG & Massive star forming region & Low mass star forming region \\
        $R_{\rm{HCN_v/HCN}}$ & $<$ 0.08 & 0.22$\pm$0.04 & 0.052$^{\rm{a, b}}$ & $<$ 0.009$^{\rm{c}}$ \\
        Linear scale [pc] & 105$\times$84 & 720 & 0.05 & 0.02 \\
       Ref. & 1 & 2 & 3 & 4 \\ \hline
    \end{tabular}
    \end{center}
    \footnotetext
    {
    {\bf{Notes.}} (a) No error was provided in the original paper. (b) The peak intensity ratio is 0.14. 
    (c) A 3 $\sigma$ upper limit of the vibrationally excited HCN 
    was estimated from noise level at near frequency. 
    {\bf{References.}} --- (1) This work; (2) Sakamoto et al. 2010; (3) Schilke et al. 1997; (4) Watanabe et al. 2012
    }
    \end{minipage}
\end{table*}

\subsection{Molecular line diagnostics: $R_{\rm{HCN/HCO^+}}$ vs. $R_{\rm{HCN/CS}}$}
Among several line emissions displayed in Figure \ref{B7spec}, 
the extremely high HCN ($J$ = 4--3) to CS ($J$ = 7--6) integrated intensity ratio 
($\equiv$ $R_{\rm{HCN/CS}}$, hereafter) of $>$ 12.7 in brightness temperature scale stands out. 
We searched the literature to check whether such ratio is commonly observed or not, and the results are listed in Table \ref{CS_table}. 
This table includes three Seyfert galaxies (NGC 1097, NGC 1068 and NGC 4418) and two nuclear starburst galaxies (NGC 253 and M 82). 
Although the physical volumes probed are different, only NGC 1097 and NGC 1068 exhibit 
high $R_{\rm{HCN/CS}}$ among these galaxies. 
This behavior is also observed in lower-$J$ lines. 
For example, the HCN ($J$ = 1--0) to CS ($J$ = 2--1) line ratios of 
NGC 1068 and NGC 253 obtained by a recent NRO 45m single dish observation 
(Nakajima et al. 2011) show a significant difference, i.e., $\sim$ 5.5 in NGC 1068 and $\sim$ 1.8 in NGC 253. 
A similar high value as seen in NGC 1068 is also found in NGC 1097 
by our ALMA band 3 observation ($\sim$ 3.7; Kohno et al. in prep.). 
In addition, the multi-molecular work by Aladro et al. (2013), where more than a dozen of species were observed, point out the
H$^{13}$CN ($J$ = 1--0) to C$^{34}$S ($J$ = 2--1) line ratio (less affected by opacity than the low-$J$ transitions of the main isotopologues) 
as showing a large variation between the AGN and SB dominated environments. 
NGC 1068 also shows a high HCN ($J$ = 3--2) to CS ($J$ = 5--4) line ratio of 6.3 (Kamenetzky et al. 2011). 
From these results, it seems that the line ratio of HCN to CS is high if the galaxy hosts an AGN. 
Additionally, we derived the $R_{\rm{HCN/HCO^+}}$ ratio of these galaxies, 
as this ratio has been proposed as a discriminator for 
AGN and starburst activity (e.g., Kohno et al. 2001; Kohno 2005; Krips et al. 2008; Davies et al. 2012). 

\begin{figure*}
  \begin{center}
    \FigureFile(120mm,120mm){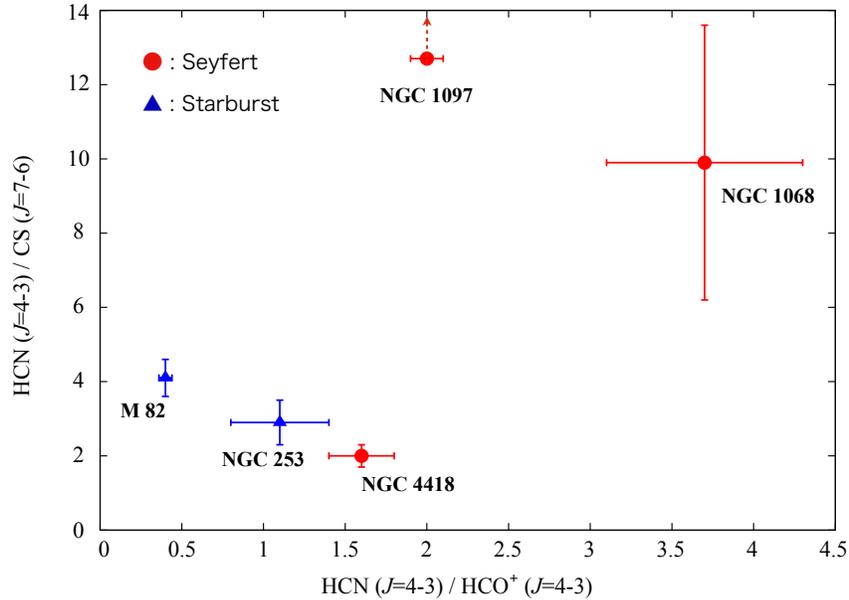}
  \end{center}
  \caption{
  HCN ($J$ = 4--3) to HCO$^+$ ($J$ = 4--3) and HCN ($J$ = 4--3) to CS ($J$ = 7--6) 
  integrated intensity ratios of the Seyfert (circles) and starburst galaxies (triangles) in brightness temperature scale . 
  Note that all three Seyfert galaxies here contain contributions of star formation 
  due to, e.g., nuclear star formation and/or the lack of angular resolution. 
  Two Seyfert galaxies, NGC 1097 and NGC 1068 exhibit enhanced HCN ($J$ = 4--3) emission 
  with respect to HCO$^+$ ($J$ = 4--3) and CS ($J$ = 7--6) lines. 
  NGC 4418 is thought to have an AGN, however, the line ratios are similar to those of starburst galaxies. 
  This galaxy shows many unusual features (e.g., Spoon et al. 2001; Gonz{\'a}lez-Alfonso et al. 2012) and the interpretation therefore is not straightforward.
    		}
  \label{diag}
\end{figure*}

Then, by using these two line ratios, we constructed a diagram, 
referred to as the $``$submm-HCN diagram$"$ hereafter, displayed in Figure \ref{diag}. 
To avoid confusion with the previous $``$HCN diagram$"$ proposed by Kohno et al. (2001), we refer to that as the $``$mm-HCN diagram$"$. 
Note that one advantage of this submm-HCN diagram over the mm-HCN diagram is that 
it is much more applicable to high redshift galaxies because its lines are 
within the frequency coverage of ALMA up to redshift $z$ $\sim$ 3. 
At first inspection, we find that both starburst galaxies are located in the bottom-left region in the diagram. 
On the other hand, the two AGN-host galaxies, NGC 1097 and NGC 1068, are located in the top-right region, 
although NGC 1097 has a compact nuclear star-forming region (Prieto et al. 2005; Davies et al. 2007) 
and the low angular resolutions of the NGC 1068 observations (Bayet et al. 2009; P\'{e}rez-Beaupuits et al. 2008) 
are likely contaminated by the circumnuclear starburst ring (the case of NGC 4418 is discussed separately later). 
From the diagram, one might deduce that a galaxy where an AGN contributes resides in the top-right region. 
However, this is a tentative view as 
we do not have clear theoretical evidence to support the use of this diagram, and it is based on few samples. 
Clearly, we need large samples and theoretical models to test extensively this diagram.

To get an insight on why AGN host galaxies seem to exhibit high line ratios in this diagram, 
let us first consider the excitation conditions only. 
Considering the critical density 
($n_{\rm crit}$ = 8.5 $\times$ 10$^6$ cm$^{-3}$ for HCN ($J$ = 4--3), 1.8 $\times$ 10$^6$ cm$^{-3}$ for HCO$^+$ ($J$ = 4--3), and 2.9 $\times$ 10$^6$ cm$^{-3}$ for CS ($J$ = 7--6), 
respectively\footnote{calculated for the kinetic temperature $T_k$ = 100 K in the optically thin limit.}; Greve et al. 2009) 
and the level energy ($E_u/k_{\rm B}$, see Table \ref{line}) of each line, 
a galaxy which shows a high $R_{\rm HCN/CS}$ in Figure \ref{diag} is expected to have higher gas density and lower temperature, 
compared to those that show low $R_{\rm HCN/CS}$. 
From the same perspective, a galaxy which shows a high $R_{\rm HCN/HCO^+}$ is expected to also have higher gas density, although the temperature is not restricted tightly 
because the level energy of HCN ($J$ = 4--3) and HCO$^+$ ($J$ = 4--3) are almost the same. 

In addition to these excitation conditions, different relative molecular abundances 
and/or the effect of IR-pumping could also lead to the different line ratios seen in this diagram. 

For example, in the case of starburst galaxies, which are located in the bottom-left region, the fractional abundance of HCO$^+$ 
can be increased due to frequent supernovae explosions (more generally, high ionization effects from cosmic rays; Nguyen et al. 1992, Meijerink et al. 2006), 
which would lead to low $R_{\rm{HCN/HCO^+}}$. 
This supernovae effect can also explain a smaller $R_{\rm{HCN/HCO^+}}$ in M 82 than in NGC 253. 
As the starburst age in NGC 253 is younger than that in M 82, 
the molecular gas heating would be dominated by shocks in NGC 253 (e.g., Mart$\acute{\i}$n et al 2006), 
whereas it seems to be dominated by PDRs and/or SNe in M 82. 

In the case of NGC 1097, although the cause of the high ratios is not clear at this stage, 
at least non-collisional excitation (IR pumping) can now be rejected due to the non-detection of 
vibrationally excited HCN as shown in the previous subsection. 
A possible explanation for this galaxy could thus be different excitation conditions (gas density and temperature) 
and/or enhanced HCN abundance with respect to CS and HCO$^+$. 
These possibilities are discussed in detail in the subsequent sections.

One may find that NGC 4418 shows line ratios comparable to those in nuclear starburst galaxies, although it has a type-2 AGN. 
A similar phenomenon has been found in the mm-HCN diagrams, 
i.e., some Seyfert galaxies exhibit low $R_{\rm{HCN/HCO^+}}$ and $R_{\rm{HCN/CO}}$ ratios 
which are comparable to starburst galaxies (e.g., Kohno 2005). 
In the case of NGC 4418, although the PAH features indicate that the starburst is associated with the nucleus (e.g., Imanishi et al. 2004), 
the energy budget in the nuclear region of NGC 4418 seems to be dominated by a buried AGN 
based on the AGN-like IR spectrum (e.g., Spoon et al. 2001) and the high HCN ($J$ = 1--0) to HCO$^+$ ($J$ = 1--0) line ratio (Imanishi et al. 2004). 
Therefore, the low $R_{\rm HCN/CS}$ is apparently inconsistent with the case of the other Seyferts proposed in the diagram. 
One possible explanation for this situation would be IR-pumping, 
which could significantly affect the molecular rotational populations, 
and thus we are not measuring the ratios of pure rotational lines. 
Note that a clear absorption feature of HCN at 14.0 $\mu$m is detected (Lahuis et al. 2007; Veilleux et al. 2009), 
and the vibrationally excited line is also observed (Sakamoto et al. 2010) in NGC 4418. 

\begin{table*}[h]
  \caption{HCN($J$ = 4--3) / CS($J$ = 7--6) and HCN ($J$ = 4--3) / HCO$^+$ ($J$ = 4--3) 
  integrated intensity ratios in brightness temperature scale.}
  \label{CS_table}
  \begin{minipage}{\textwidth}
  \begin{center}
    \begin{tabular}{cccccc}
    \hline\hline
        & NGC 1097(nucleus) & NGC 1068 & NGC 4418 & NGC 253 & M82 \\ \hline
        Type & LLAGN & AGN & LIRG & SB & SB \\
        $R_{\rm{HCN/CS}}$ & $>$12.7 & 9.9$\pm$3.7 &  2.0$\pm$0.3 & 2.9$\pm$0.6 & 4.1$\pm$0.5 \\
        $R_{\rm{HCN/HCO^+}}$ & 2.0$\pm$0.2 & 3.7$\pm$0.6& 1.6$\pm$0.2 & 1.1$\pm$0.3 & 0.4$\pm$0.04 \\
        Linear scale [pc] & 105$\times$84 & 1060 & 720 & 290 & 300 \\ 
        Ref. & 1 & 2 & 3 & 4 & 5 \\ \hline
    \end{tabular}
    \end{center}
    \footnotetext
    {In the case of NGC 1097, 10 \% absolute flux error is assumed. 
    {\bf{References.}} --- (1) This work; (2) Bayet et al. 2009; P${\acute{\rm{e}}}$rez-Beaupuits et al. 2009; 
    (3) Sakamoto et al. 2010; (4) Bayet et al. 2009; Knudsen et al. 2007 (5) Bayet et al. 2008; Seaquist et al. 2000
    }
    \end{minipage}
\end{table*}

\section{Physical conditions of the dense gas}
In this section, we investigate the physical condition (gas density, excitation temperature, and abundance) of HCN and HCO$^+$, 
under both LTE and non-LTE conditions. 

\subsection{Column density under LTE}
Assuming optically thin and local thermodynamic equilibrium (LTE) conditions, 
we estimate the rotational temperatures ($T_{\rm{rot}}$) and column densities ($N_{\rm{mol}}$) 
of HCN and HCO$^+$ molecules in the nuclear region via excitation diagrams (rotation diagrams). 
An excitation diagram is a plot of the column density per statistical weight of a number of molecular energy levels, 
as a function of their energy above the ground state. 
From the optically thin condition, the column density of the level $u$ ($N_u$) can be written as 
\begin{equation}
N_u = \frac{8\pi k_{\rm{B}} \nu^2 \int{T_{\rm{b}}}dv}{hc^3 A_{ul}}
\end{equation}
where $k_{\rm{B}}$ and $h$ are the Boltzmann and Planck constants, $A_{ul}$ is an Einstein A coefficient from upper to lower state, 
$\nu$ the frequency of the line, $c$ the speed of light, and $T_{\rm b}$ the brightness temperature of the line. 
Also from the LTE condition, $N_u$ is written as 
\begin{equation}
N_u = \frac{N_{\rm{mol}}}{Q(T)} g_u \exp \Biggl(-\frac{E_u}{k_{\rm B}} T_{\rm{ex}}\Biggr)
\end{equation}
where $N_{\rm{mol}}$ is the total column density of a given molecule, $Q(T)$ is a partition function, 
and $E_u$ is an energy at level $u$ above the ground state. 
Therefore, the decimal logarithm of $N_u/g_u$ versus $E_u/k_{\rm B}$ will yield a straight line with a slope of $-\log{e}/T_{\rm{ex}}$. 
This excitation temperature is called the ``rotational temperature'' ($T_{\rm{rot}}$) simply because only rotational transitions are measured observationally here. 
Column densities can also be derived from the intercept of the y-axis in the excitation diagram. 
In fact, assuming LTE implies that $T_{\rm ex}$ is the same for all transitions and equal to $T_{\rm rot}$. 
On the other hand, because the gas may not be thermalized, 
the derived $T_{\rm{rot}}$ is thus only a lower limit to the kinetic temperature, $T_{\rm{kin}}$. 
The detailed method is described in Goldsmith \& Langer (1999). 

\begin{figure*}
  \begin{center}
    \FigureFile(180mm,180mm){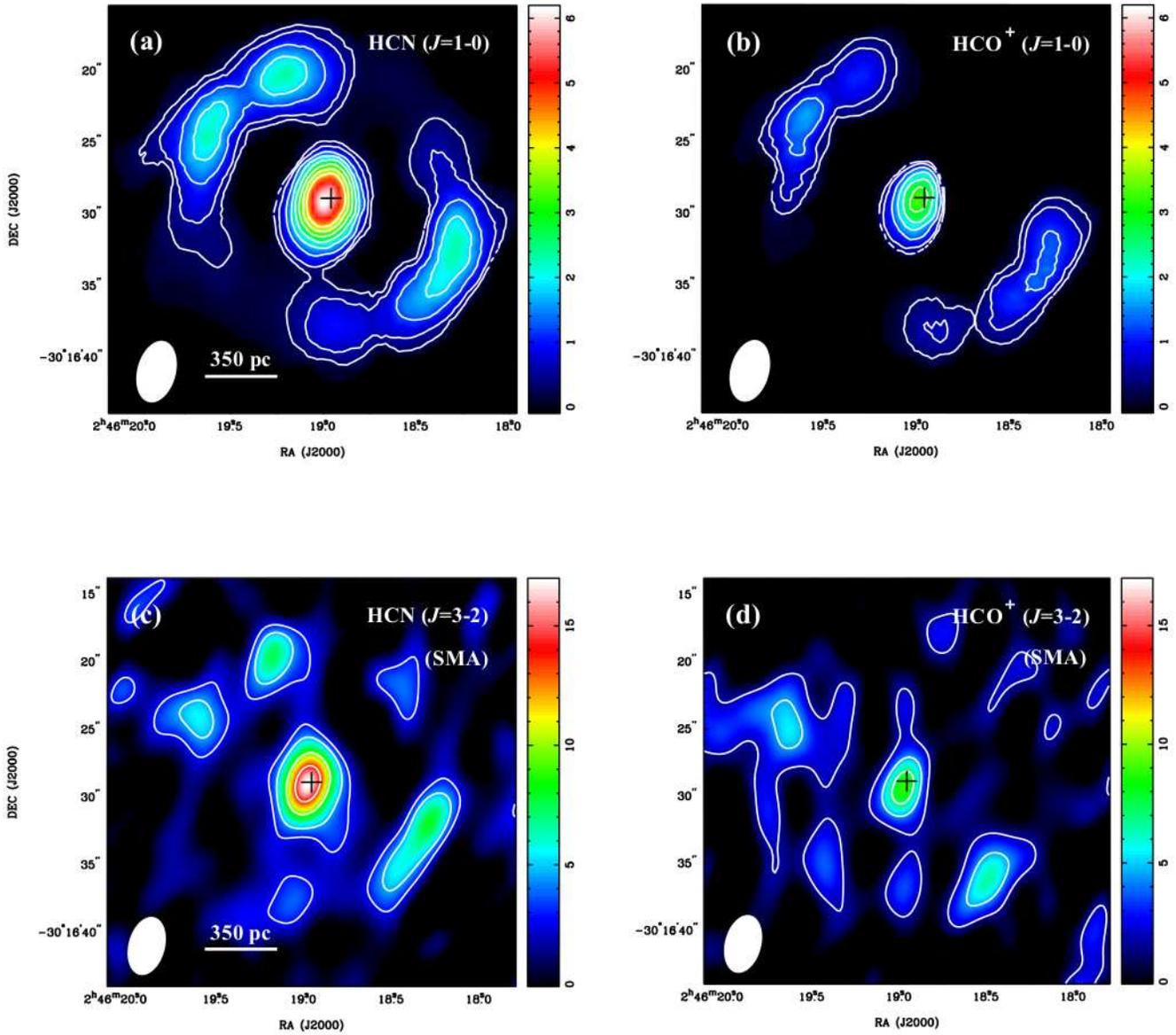}
  \end{center}
  \caption{
  Integrated intensity maps in the central 30$''$ $\times$ 30$''$ (2.1 kpc $\times$ 2.1 kpc) region of NGC 1097, 
  derived by calculating the zeroth moment of ALMA (for $J$ = 1--0: from Kohno et al. in prep.) 
  and SMA (for $J$ = 3--2: from Hsieh et al. 2012) data cubes. 
  Although we focus on the nucleus in this paper, the circumnuclear starburst ring is also clearly detected. 
  Central crosses indicate the peak positions of the 860 $\mu$m continuum 
  and the color bars are in units of Jy beam$^{-1}$ km s$^{-1}$. 
  All beam sizes are convolved to 4$''$.4 $\times$ 2$''$.7, P.A. = -14.6$^\circ$ and shown in the bottom left corners. 
  (a) Integrated intensity map of the HCN ($J$ = 1--0) emission. 
  The value at the 860 $\mu$m peak position is 5.99 Jy beam$^{-1}$ km s$^{-1}$. 
  The contour levels are 5, 10, 20, ${}_{\cdots}$, and 70 $\sigma$, where 1 $\sigma$ is 0.06 Jy beam$^{-1}$ km s$^{-1}$. 
  (b) Integrated intensity map of the HCO$^+$ ($J$ = 1--0) emission. 
  The value at the 860 $\mu$m peak position is 3.18 Jy beam$^{-1}$ km s$^{-1}$. 
  The contour levels are 5, 10, 20, ${}_{\cdots}$, and 50 $\sigma$, where 1 $\sigma$ is 0.06 Jy beam$^{-1}$ km s$^{-1}$. 
  (c) Integrated intensity map of the HCN ($J$ = 3--2) emission. 
  The value at the 860 $\mu$m peak position is 16.70 Jy beam$^{-1}$ km s$^{-1}$. 
  The contour levels are 3, 5, 10, and 15 $\sigma$, where 1 $\sigma$ is 0.95 Jy beam$^{-1}$ km s$^{-1}$. 
  (d) Integrated intensity map of the HCO$^+$ ($J$ = 3--2) emission. 
  The value at the 860 $\mu$m peak position is 8.45 Jy beam$^{-1}$ km s$^{-1}$. 
  The contour levels are 2, 5, and 8 $\sigma$, where 1 $\sigma$ is 0.90 Jy beam$^{-1}$ km s$^{-1}$. 
  		}
  \label{rotmom}
\end{figure*}

In order to make an excitation diagram, at least two transitions of a given molecule are necessary. 
We used our ALMA band 7 data for $J$ = 4--3 transitions, band 3 data taken from Kohno et al. (in prep.) for 1--0 transitions, 
and SMA data from Hsieh et al. (2012) for 3--2 transitions. 
We convolved all maps to 4$''$.4 $\times$ 2$''$.7 with P.A. = -14.6$^\circ$ 
(the beam of the HCO$^+$ ($J$ = 3--2) data) for comparison and then measured the line ratios. 
The convolved integrated intensity images of the 1--0 and 3--2 transitions are in Figure \ref{rotmom}. 
Although the $uv$ coverage of the $J$ = 1--0, 3--2, and 4--3 transitions are, 
5.5 -- 78.5, 7.5 -- 61.5, and 14 -- 134 k$\lambda$, respectively, 
we focus on the compact ($\sim$ 1$''$ or less) nuclear source, 
thus we do not expect significant errors in the line ratios due to differently filtered out extended flux. 

\begin{figure*}
  \begin{center}
    \FigureFile(160mm,160mm){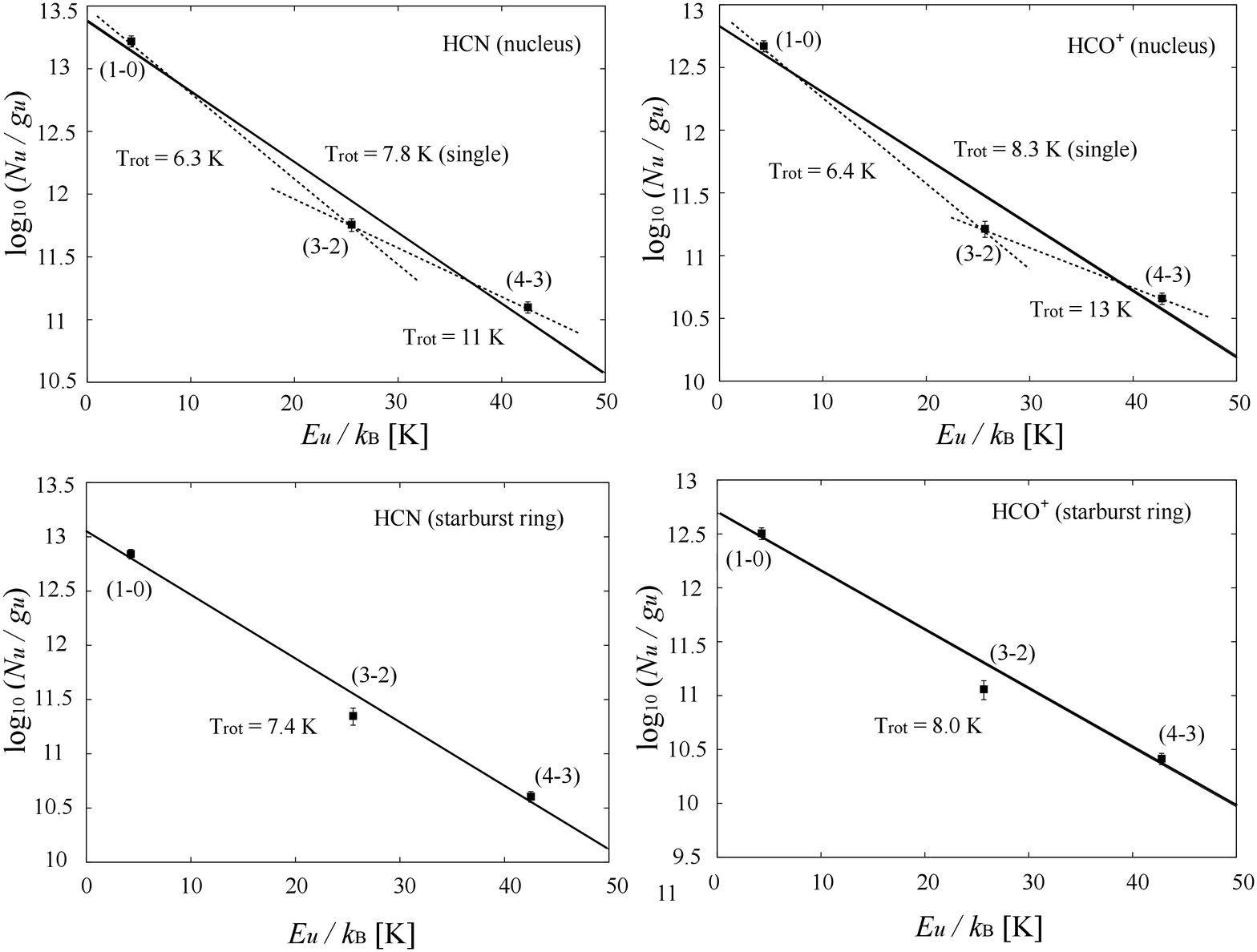}
  \end{center}
  \caption{
  Excitation diagrams of the HCN and HCO$^+$ molecules in the nucleus ({\it{top left and right}}) and starburst ring ({\it{bottom left and right}}). 
  The data of the circumnuclear SB ring are measured at GMA 3 (see details in the text). 
  We used $J$ = 1--0, 3--2 and 4--3 transitions. 
  Detections are represented by black square symbols with error bars. 
  The linear regression using a single-component is represented by solid lines, 
  while the two-components regression is represented by dashed lines.  
  The resulting rotational temperatures for both fits are overlaid. 
  		}
  \label{rot}
\end{figure*}

The parameters we used are listed in Table \ref{rottable} and 
the resulting excitation diagrams are shown in Figure \ref{rot}. 
It should be noted that these are the first ever extragalactic high resolution (100 pc scale) multi-transitional 
excitation diagrams of the dense gas in AGN environments. 
In Figure \ref{rot}, the LTE solutions are shown by solid lines, which indicate single-component fittings. 
At first, we estimated the rotational temperature and the column density of HCN and HCO$^+$ in the nucleus 
from these single-component fittings, as listed in Table \ref{rotresult}, where 
$T_{\rm{rot}}$ = 7.8$\pm$0.3 K and $N_{\rm{HCN}}$ = (2.4$\pm$0.3) $\times$ 10$^{13}$ cm$^{-2}$ for HCN, 
and $T_{\rm{rot}}$ = 8.3$\pm$0.3 K and $N_{\rm{HCO^+}}$ = (6.8$\pm$0.9) $\times$ 10$^{12}$ cm$^{-2}$ for HCO$^+$, respectively. 
Therefore, the HCN to HCO$^+$ column density ratio $N_{\rm{HCN/HCO^+}}$ is estimated as 3.5$\pm$0.6. 
This ratio indicates HCN is more abundant than HCO$^+$ in the nucleus of NGC 1097. 

It should be noted, however, that the plotted data seems not well fitted by single-component fittings as seen in Figure \ref{rot}. 
This is understandable as the existing gas is not uniform. 
Thus, we adopted a very simple approach assuming there exist two gas components in the observed region: 
one is responsible for the low-$J$ ($J$ $\leq$ 3) emission and the other for the high-$J$ ($J$ $\geq$ 3) emission. 
The results of these two-components fittings are also shown in Figure \ref{rot} by dashed lines. 
The derived $T_{\rm{rot}}$ for low/high-$J$ components are, 
6.3$\pm$0.3 K and 11.2$\pm$1.2 K for HCN, and 6.4$\pm$0.4 K and 13.4$\pm$2.1 K for HCO$^+$, respectively. 
Considering the small error bars, our assumption of at least two temperature components would be valid. 
The column density ratio $N_{\rm{HCN/HCO^+}}$ is estimated to be 3.5$\pm$0.7 by using low-$J$ components only, 
and 5.0$\pm$2.7 by using high-$J$ components only. 
Both cases indicate that HCN is more abundant than HCO$^+$. 

In order to compare this ratio in different environments, 
we also made excitation diagrams of HCN and HCO$^+$ 
in one position in the circumnuclear starburst ring of NGC 1097 (giant molecular cloud association (GMA) 3 described in Hsieh et al. 2012; 
this case is also displayed in Figure \ref{rot}, bottom) 
and the nucleus of NGC 253 (nearby starburst galaxy, Knudsen et al. 2007; Paglione et al. 1997). 
The values for the central region of NGC 1068 (nearby typical type-2 Seyfert) are taken from Aladro et al. (2013). 
Then, we derived $N_{\rm{HCN/HCO^+}}$ for these galaxies, which are listed in Table \ref{rotresult}. 
From this Table, it seems that the HCN-to-HCO$^+$ abundance ratio 
would be enhanced in the AGN-host galaxies compared to pure starburst regions/galaxies, although the statistical significance is not yet sufficient. 

Although we fitted the excitation diagrams of NGC 1097 with only two temperature components, 
it is clear that there is a contiguous change in temperature. 
Another possible explanation that can explain the low $J$ = 3--2 points on the excitation diagrams is 
that the $J$ = 3--2 lines have modest opacity (opacity peaks at the $J$ = 3--2), though, 
our non-LTE analysis in the next subsection implies that at least high-$J$ lines in the excitation diagrams are optically thin. 
Our hypothesis that there are at least two temperature components in the nucleus of NGC 1097, 
is further investigated in the next subsection. 

\begin{table*}
  \caption{Line parameters used to make excitation diagrams}
  \label{rottable}
  \begin{minipage}{\textwidth}
  \begin{center}
    \begin{tabular}{cccccccc}
    \hline\hline
       Emission & $\nu_{\rm{rest}}$ & $E_u / k_{\rm B}$ & $A_{ul}$ &  $I$ (nucleus) & $I$ (ring) & $I$ (NGC 253)\\
         & [GHz] & [K] & [s$^{-1}$] & [K km s$^{-1}$] & [K km s$^{-1}$] & [K km s$^{-1}$]\\ 
         (1) & (2) & (3) & (4) & (5) & (6) & (7) \\ \hline
         HCN ($J$ = 1--0) & 88.631 & 4.25 & 2.4075$\times$10$^{-5}$ & 78.43$\pm$7.88 & 32.87$\pm$3.27 & 44.38\\
         HCN ($J$ = 3--2) & 265.886 & 25.52 & 8.3559$\times$10$^{-4}$ & 24.54$\pm$2.80 & 9.50$\pm$1.70 & 36.3$\pm$3.1 \\
         HCN ($J$ = 4--3) & 354.505 & 42.53 & 2.0540$\times$10$^{-3}$ & 9.49$\pm$0.96 & 3.05$\pm$0.34 & 21.79 \\
          & & & & & & & \\
         HCO$^+$ ($J$ = 1--0) & 89.188 & 4.28 & 4.5212$\times$10$^{-5}$ & 41.12$\pm$4.19 & 28.06$\pm$3.49 & 42.78 \\
         HCO$^+$ ($J$ = 3--2) & 267.558 & 25.68 & 1.4757$\times$10$^{-3}$ & 12.50$\pm$1.77 & 8.50$\pm$1.70 & - \\
         HCO$^+$ ($J$ = 4--3) & 356.734 & 42.80 & 3.6269$\times$10$^{-3}$ & 6.01$\pm$0.62 & 3.43$\pm$0.41 & 22.70 \\ \hline
    \end{tabular}
    \footnotetext{{\bf{Notes.}} Integrated intensities and line luminosities were measured by a 4$''$.4 $\times$ 2$''$.7, P.A. = -14.6$^\circ$ beam. 
    In the case of NGC 1097, a 10 \% absolute flux error is assumed for all transitions, 
    and also the primary beam attenuation has been corrected. 
    Column 1: Line name. Columns 2, 3, 4: Rest frequency, upper energy level, and 
    Einstein A coefficient, respectively. These values are cited from LAMDA (Sch\"{o}ier et al. 2005). 
    Columns 5, 6, 7: Velocity integrated intensity of the nucleus (860 $\mu$m peak) and 
    a giant molecular cloud in the starburst ring (GMA 3 in Hsieh et al. (2012))
    by calculating the zeroth moment of each data cube. 
    We used channels which contain $>$ 3 $\sigma$ emission for $J$ = 1--0 and 4--3 transitions. 
    For $J$ = 3--2 transitions in the nucleus, we fixed the velocity range to be integrated because the S/N ratios are relatively low. 
    For $J$ = 3--2 transitions in GMA 3, we used the integrated intensities in Hsieh et al. (2012). 
    {\bf{References.}} NGC 1097: $J$ = 1--0, 3--2, and 4--3 data are from Kohno et al. in prep., Hsieh et al. (2012), and this work, respectively. 
    NGC 253: $J$ = 1--0 and 4--3 transitions are from Knudsen et al. (2007), 3--2 is from Paglione, Jackson, Ishizuki (1997).}
    \end{center}
    \end{minipage}
\end{table*}

\begin{table*}
  \caption{Resulting parameters from excitation diagrams}
  \label{rotresult}
  \begin{minipage}{\textwidth}
  \begin{center}
    \begin{tabular}{cccccc}
    \hline\hline
       & NGC 1097 & NGC 1097 & NGC 253 & NGC 1068 \\
       & (nucleus) & (ring) &  & & \\ \hline
       $T_{\rm{rot}}$(HCN) [K] & 7.76$\pm$0.26 & 7.40$\pm$0.23 & 11.0$\pm$1.7 & 5.0$\pm$0.3 \\
       $T_{\rm{rot}}$(HCO$^+$) [K] & 8.25$\pm$0.29 & 7.96$\pm$0.27 & 11.51 & 4.9$\pm$0.1 \\
       $N_{\rm{HCN}}$ [cm$^{-2}$] & (2.36$\pm$0.29)$\times$10$^{13}$ & (1.13$\pm$0.14)$\times$10$^{13}$ & 
       (1.18$\pm$0.50)$\times$10$^{13}$ & (1.4$\pm$0.2)$\times$10$^{15}$ \\
       $N_{\rm{HCO^+}}$ [cm$^{-2}$] & (6.84$\pm$0.85)$\times$10$^{12}$ & (4.98$\pm$0.62)$\times$10$^{12}$ & 7.08$\times$10$^{12}$ & (5.3$\pm$0.2)$\times$10$^{14}$ \\ 
       $N_{\rm{HCN}}/N_{\rm{HCO^+}}$ & 3.45$\pm$0.60 & 2.27$\pm$0.40 & 1.67$\pm$0.71 & 2.64$\pm$0.39 \\ 
       Beamsize [arcsec] & 4$''$.4$\times$2$''$.7 & 4$''$.4$\times$2$''$.7 & 27$''$ & 28$''$ \\
       Ref. & 1 & 2 & 3 & 4 \\ \hline 
          \end{tabular}
    \footnotetext{{\bf{References.}} (1), (2): This work, Hsieh et al. 2012, Kohno et al. in prep., (3): Knudsen et al. 2007, Paglione, Jackson, Ishizuki 1997, (4): Aladro et al. 2013.}
    \end{center}
    \end{minipage}
\end{table*}

\subsection{Non-LTE excitation analysis in the nucleus}
We also ran non-LTE models via RADEX (van der Tak et al. 2007) for a spherical geometry 
to constrain the excitation conditions ($T_{\rm{kin}}$, $n_{\rm{H_2}}$ and $N_{\rm{mol}}$) 
of the HCN and HCO$^+$ emission lines detected in the nucleus. 
RADEX uses an escape probability approximation to solve the non-LTE excitation 
in a homogeneous (single temperature, single density), single component medium, i.e., all HCN and HCO$^+$ transitions 
are assumed to be emitted from the same region. 
Again, we used the $J$ = 4--3, 3--2 and 1--0 transitions for this analysis. 
All line specifications (spectroscopic data and collisional excitation rates) were taken from the Leiden Atomic and Molecular Database (LAMDA; Sch\"{o}ier et al. 2005). 
For our non-LTE analysis, we varied\footnote{We used a script developed by Y. Tamura for this calculation: http:\slash\slash{}www.ioa.s.u-tokyo.ac.jp\slash{}{\%}7Eytamura
\slash{}Wiki\slash{}?Science{\%}2FUsingRADEX}
the gas kinetic temperature within a range of $T_{\rm{kin}}$ = 10 -- 600 K 
using steps of $dT_{\rm{kin}}$ = +10 K, and a gas density of $n_{\rm{H_2}}$ = 10$^2$ -- 10$^7$ cm$^{-3}$ 
using steps of $dn_{\rm{H_2}}$ = $\times$ 10$^{0.25}$. 
In the following, a relative abundance ratio of a molecule $X$ to another molecule $Y$ is expressed as [$X$]/[$Y$]. 

In our RADEX simulation, we 
\begin{itemize}
\item changed the H$_2$ column density, $N_{\rm H_2}$, from 1.0 $\times$ 10$^{20}$ cm$^{-2}$ to 1.0 $\times$ 10$^{25}$ cm$^{-2}$. 
\item fixed [HCO$^+$]/[H$_2$] = 5.0 $\times$ 10$^{-9}$, the standard value observed in Galactic molecular clouds (e.g., Blake et al. 1987), 
therefore we changed $N_{\rm {HCO^+}}$ from 5.0 $\times$ 10$^{11}$ cm$^{-2}$ to 5.0 $\times$ 10$^{16}$ cm$^{-2}$.
\item changed [HCN]/[HCO$^+$] to 1, 2, 3,$\ldots$, 15, 20, 30, 40, 50, $\ldots$, and 100 in each $N_{\rm {HCO^+}}$ case. 
\item fixed the line velocity width to 250 km s$^{-1}$ for all cases, which is roughly a mean value of the FWZIs of HCN ($J$ = 4--3) and HCO$^+$ ($J$ = 4--3) shown in Section 4. 
\end{itemize}
Therefore, we tested wide range of a ratio of gas density to velocity gradient, or equivalently gas column density to line velocity width, 
$N_{\rm HCO^+}$/$dV$ [cm$^{-2}$ (km s$^{-1}$)$^{-1}$] from 2.0 $\times$ 10$^{9}$ to 2.0 $\times$ 10$^{14}$. 
The $T_{\rm{bg}}$ = 2.73 K background emission was also added to the calculation. 

Under these conditions, we ran RADEX for each ($N_{\rm {HCO^+}}$/$dV$, [HCN]/[HCO$^+$]) case by changing $n_{\rm H_2}$ and $T_{\rm {kin}}$, 
and carried out a $\chi^2$ test to search for the best parameters to explain the observed intensities. 
Then we confirmed that when 1.0 $\times$ 10$^{20}$ cm$^{-2}$ $\leq$ $N_{\rm H_2}$ $\leq$ 1.0 $\times$ 10$^{24}$ cm$^{-2}$, 
all cases yielded very similar trends with relatively small $\chi^2$ values ($<$ 0.1). 
Otherwise, when $N_{\rm H_2}$ $>$ 1.0 $\times$ 10$^{24}$ cm$^{-2}$, we could not achieve reasonable solutions based on our $\chi^2$ tests. 
Therefore, we present here the case of $N_{\rm H_2}$ = 4.8 $\times$ 10$^{22}$ cm$^{-2}$ as a representation. 
This $N_{\rm H_2}$ value is derived from the CO ($J$ = 1--0) integrated intensity by Kohno et al. (2003) with a CO-to-H$_2$ conversion factor 
$X_{\rm CO}$ = 5.0 $\times$ 10$^{20}$ cm$^{-2}$ (K km s$^{-1}$)$^{-1}$ (Pi{\~n}ol-Ferrer et al. 2011).

Following are the results of our $\chi^2$ tests. 
All line ratios we used are listed in Table \ref{chisqparams}. 

\begin{figure*}
  \begin{center}
    \FigureFile(180mm,180mm){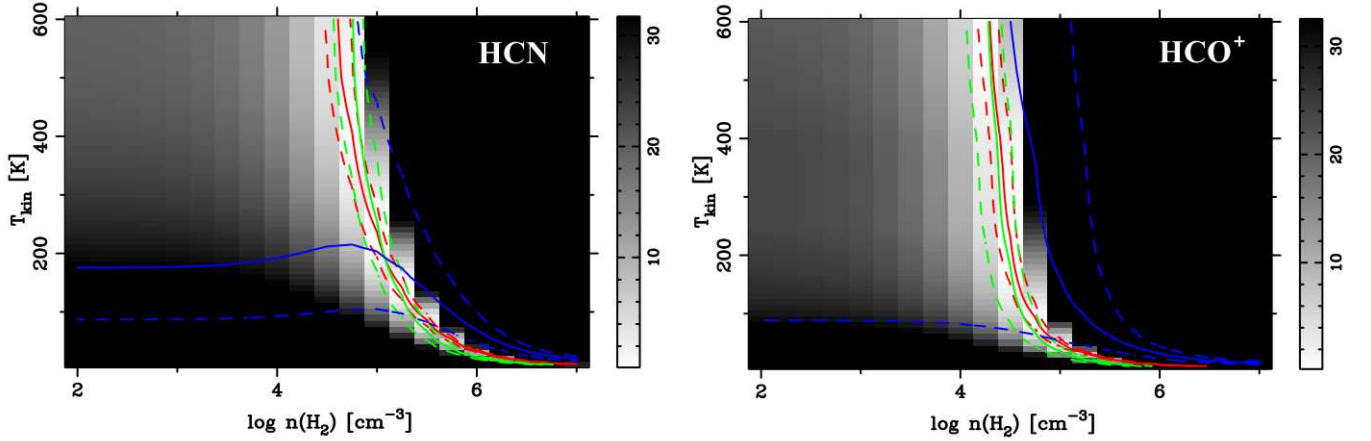}
  \end{center}
  \caption{
 The $\chi^2$ test results of the non-LTE analysis carried out with RADEX (see text for explanation) 
 based on HCN ({\it{left}}) and HCO$^+$ ({\it{right}}) data for different kinetic temperatures and gas densities. 
 The molecular hydrogen column density and the relative abundance ratio of these molecules to H$_2$ 
 were fixed to $N_{\rm{H_2}}$ = 4.8 $\times$ 10$^{22}$ cm$^{-2}$ 
 and [HCN]/[H$_2$] = [HCO$^+$]/[H$_2$] = 5.0 $\times$ 10$^{-9}$. 
 The red, green and blue lines are tracks of $R^{\rm{mol}}_{43/10}$, $R^{\rm{mol}}_{32/10}$ and $R^{\rm{mol}}_{43/32}$ ratios, respectively.
 The dashed lines are $\pm$ 1 $\sigma$ error of each ratio. 
 The background gray scale indicates the $\chi^2$ value of each point. 
 It is clear that only the $R^{\rm{mol}}_{43/32}$ tracks show trends different from the others. 
   		}
  \label{LVG_1}
\end{figure*}

\begin{figure*}
  \begin{center}
    \FigureFile(120mm,120mm){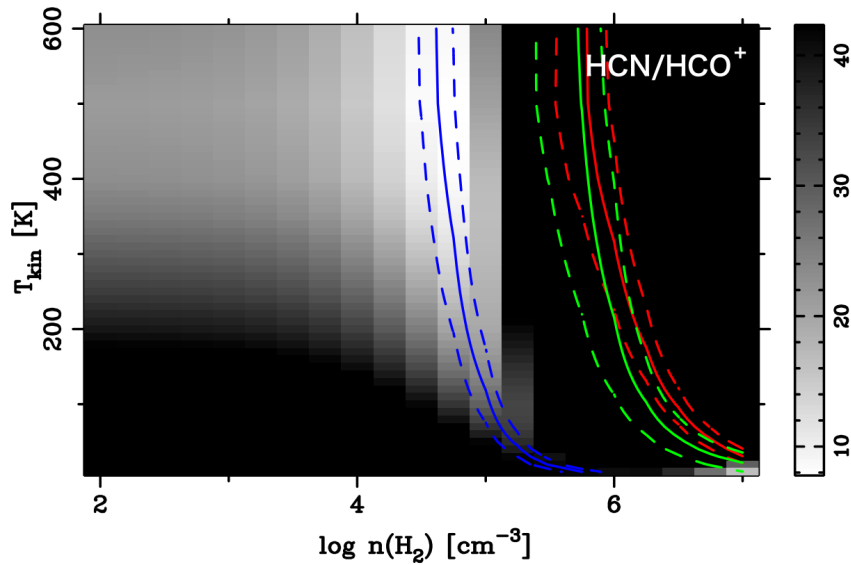}
  \end{center}
  \caption{
 The representative results of the $\chi^2$ test of the non-LTE analysis carried out with RADEX 
 to search for the best parameter set to explain the observed $R^{\rm{HCN/HCO^+}}_{43/43}$ = 1.55$\pm$0.43 (red), 
 $R^{\rm{HCN/HCO^+}}_{32/32}$ = 1.96$\pm$0.48 (green) and $R^{\rm{HCN/HCO^+}}_{10/10}$ = 1.87$\pm$0.27 (blue). 
 The molecular hydrogen column density, HCO$^+$ to H$_2$ abundance ratio, and HCN to HCO$^+$ abundance ratio were fixed to 
 $N_{\rm{H_2}}$ = 4.8 $\times$ 10$^{22}$ cm$^{-2}$, [HCO$^+$]/[H$_2$] = 5.0 $\times$ 10$^{-9}$, and [HCN]/[HCO$^+$] = 5, respectively. 
 The dashed lines are $\pm$ 1 $\sigma$ error of each track. 
 The background gray scale indicates the $\chi^2$ value of each point. 
 The best fitted parameters achieved by using these three tracks are 
 ($n_{\rm{H_2}}$, $T_{\rm{kin}}$, $\chi^2$) = (10$^{4.75}$ cm$^{-3}$, 500 K, 7.774). 
   		}
  \label{LVG_2}
\end{figure*}

First, we carried out a $\chi^2$ test including constraints of 
$R^{\rm{HCN}}_{43/10}$, $R^{\rm{HCN}}_{32/10}$, $R^{\rm{HCN}}_{43/32}$ to investigate the excitation condition of each molecule. 
Here, the notation of a molecular integrated intensity ratio is defined in a similar way to Krips et al. (2008) as
\begin{equation}
R^{\rm{mol}}_{J_u, J_l/J'_u, J'_l} \equiv \frac{I^{\rm{mol}}_{J_u, J_l}}{I^{\rm{mol}}_{J'_u, J'_l}},
\end{equation}
where $J_u$ and $J_l$ indicate upper and lower rotational quantum number, respectively, 
and $I^{\rm{mol}}_{J_u, J_l}$ is the integrated intensity of 
a given molecule in erg  s$^{-1}$ cm$^{-2}$ unit at transition ($J$ = $J_u$--$J_l$). 
Observed values for these lines are listed in Table \ref{rottable}. 
Then, we repeated the test based on $R^{\rm{HCO^+}}_{43/10}$, $R^{\rm{HCO^+}}_{32/10}$, 
and $R^{\rm{HCO^+}}_{43/32}$ in the same manner. 
The typical results of this analysis are shown in Figure \ref{LVG_1}. 
It is clear that $R^{\rm{mol}}_{43/10}$ and $R^{\rm{mol}}_{32/10}$ show almost the same trend, 
whereas $R^{\rm{mol}}_{43/32}$ behaves differently in both plots of Figure \ref{LVG_1}. 
As one reasonable hypothesis, we can assume that the low-$J$ emission and high-$J$ emission are emitted from gas with 
different excitation conditions, i.e., they are emitted from different regions within the observing beam. 
Note that this is the same hypothesis as the one derived from the rotation diagrams in the previous subsection. 
Figure \ref{LVG_2}, which further reinforces this hypothesis, is an example of the $\chi^2$ test results 
based on $R^{\rm{HCN/HCO^+}}_{43/43}$, $R^{\rm{HCN/HCO^+}}_{32/32}$, $R^{\rm{HCN/HCO^+}}_{10/10}$, 
and [HCN]/[HCO$^+$] = 5 constraints. 
Here, the HCN to HCO$^+$ line ratio is defined as
\begin{equation}
R^{\rm{HCN/HCO^+}}_{J_u, J_l/J'_u, J'_l} \equiv \frac{I^{\rm{HCN}}_{J_u, J_l}}{I^{\rm{HCO^+}}_{J'_u, J'_l}}, 
\end{equation} 
and the plotted red, green, blue tracks indicate $R^{\rm{HCN/HCO^+}}_{43/43}$, $R^{\rm{HCN/HCO^+}}_{32/32}$, and $R^{\rm{HCN/HCO^+}}_{10/10}$, respectively. 
Although we could achieve the best fit parameters for these three tracks locally, 
again only the track of $J$ = 1--0 transitions exhibits a different trend, 
and the best fitted parameter set for these three tracks is 
($T_{\rm{kin}}$ = 500 K, $n_{\rm{H_2}}$ = 10$^{4.75}$ cm$^{-3}$, $\chi^2$ = 7.774), 
from which we consider the $\chi^2$ value is too high. 
These trends persist whichever [HCN]/[HCO$^+$] we adopted, and are not plausible. 
Therefore, we conclude that the low-$J$ and high-$J$ transitions are from different regions, 
thus we use only the high-$J$ (4--3 and 3--2) transitions hereafter because these transitions exhibit similar trends. 
The $J$ = 1--0 transitions can not strongly restrict the excitation conditions by themselves, thus the results are not shown in this paper. 
 
\begin{figure*}
  \begin{center}
    \FigureFile(130mm,130mm){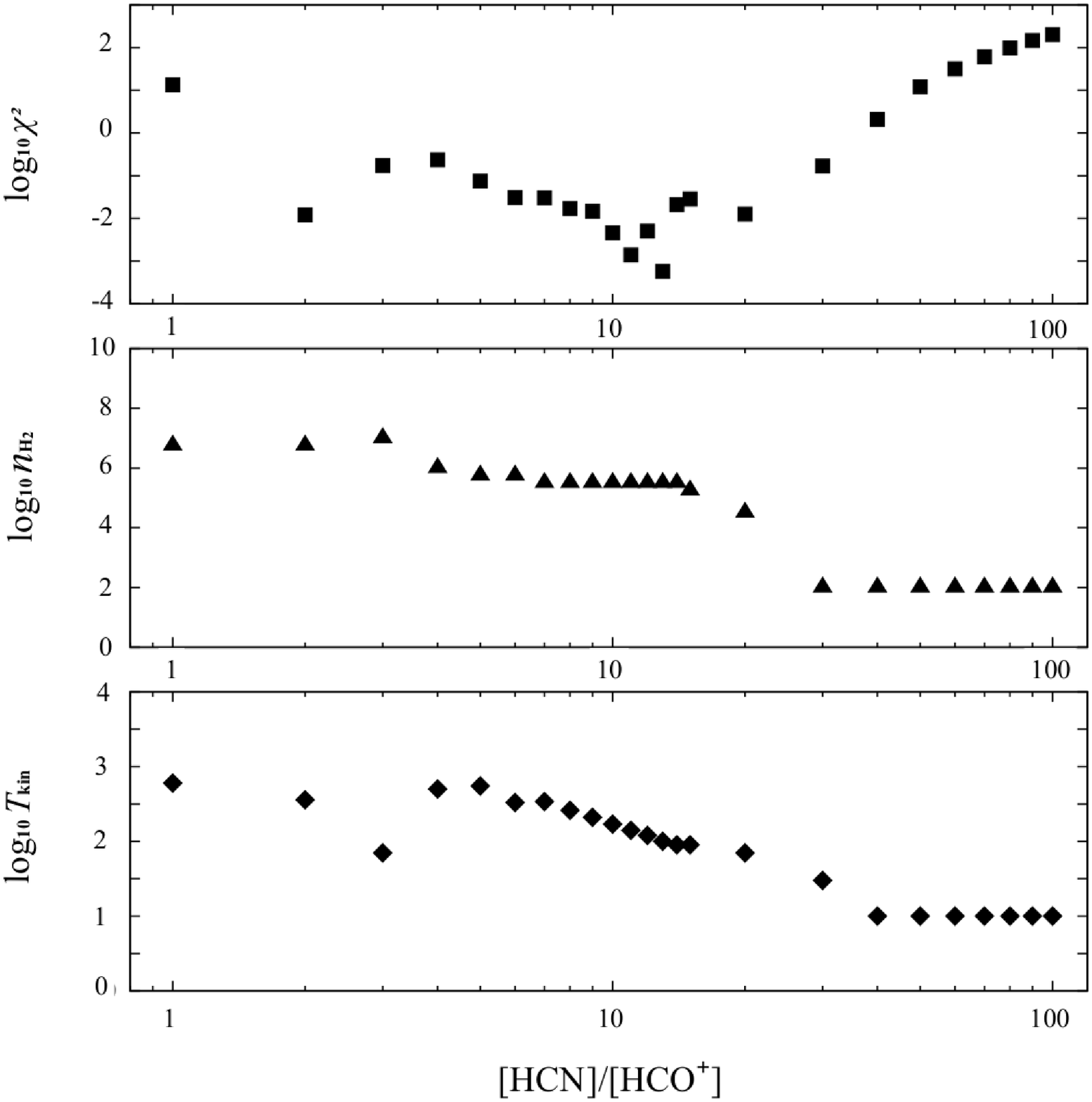}
  \end{center}
  \caption{
 The $\chi^2$ test results based on $R^{\rm{HCN/HCO^+}}_{43/43}$ and $R^{\rm{HCN/HCO^+}}_{32/32}$ constraints. 
 We changed the [HCN]/[HCO$^+$] abundance ratio from 1 to 100, while the [HCO$^+$]/[H$_2$] abundance ratio was fixed to 5.0 $\times$ 10$^{-9}$. 
 For each [HCN]/[HCO$^+$] case, we searched the best fitted point (i.e., the point with the minimum $\chi^2$ value) in the parameter plane such as shown in Figure 14 and 15. 
 The resultant values of $\chi^2$ ({\it{top}}), $n_{\rm H_2}$ ({\it{middle}}), and $T_{\rm {kin}}$ ({\it{bottom}}) of the point are plotted as functions of [HCN]/[HCO$^+$].
 Within the range of well fitted cases ($\chi^2$ $\leq$ 0.1), both $n_{\rm{H_2}}$ and $T_{\rm{kin}}$ 
 show a small decrease as [HCN]/[HCO$^+$] becomes larger. 
   		}
  \label{chi2}
\end{figure*}

 We thus changed [HCN]/[HCO$^+$] again ([HCN]/[HCO$^+$] = 1, 2, 3, $\ldots$, 15, 20, 30, 40, 50, $\ldots$, and 100) 
 and searched for the best fit parameter sets 
 by a $\chi^2$ test based on $R^{\rm{HCN/HCO^+}}_{43/43}$ and $R^{\rm{HCN/HCO^+}}_{32/32}$ constraints. 
 The resultant $\chi^2$ values of the best fitted points and the excitation parameters ($n_{\rm{H_2}}$ and $T_{\rm{kin}}$) 
 of those points are displayed in Figure \ref{chi2}. 
 Because we use only two tracks for the $\chi^2$ test, the unique solution 
 of the fitting (with lowest $\chi^2$) could be achieved in most cases presented here.  
 Within the range of well fitted cases (we define as $\chi^2$ $\leq$ 0.1), 
 both $n_{\rm{H_2}}$ and $T_{\rm{kin}}$ show a small decrease when [HCN]/[HCO$^+$] becomes larger. 
 Note that the well-fitted parameters with $\chi^2$ $\leq$ 0.1 could be achieved 
 only when 5 $\leq$ [HCN]/[HCO$^+$] $\leq$ 20 \footnote{The best fit parameter sets in the cases of [HCN]/[HCO$^+$] = 2 indicates 
 a maser, i.e., the derived excitation temperatures are found to be negative.}, 
 which indicates the HCN abundance is enhanced with respect to HCO$^+$. 
 This is the same indication derived from our rotation diagram. 
 
As a whole, we conclude from Figure \ref{chi2} that the high-$J$ HCN and HCO$^+$ emission comes from 
hot (70 K $\leq$ $T_{\rm{kin}}$ $\leq$ 550 K) and dense (10$^{4.5}$ cm$^{-3}$ $\leq$ $n_{\rm{H_2}}$ $\leq$ 10$^6$ cm$^{-3}$) molecular clouds, 
where the [HCN]/[HCO$^+$] ratio is enhanced to 4 -- 20. 
This high abundance ratio is roughly consistent with the conclusion in Yamada, Wada, Tomisaka (2007) that the HCN abundance 
must be an order of magnitude higher than that of HCO$^+$ to account for the observed high $R_{\rm HCN/HCO^+}$, 
based on their hydrodynamic simulation with three-dimensional, non-LTE radiative transfer calculations. 
Considering the low $\chi^2$ values, 
these fitting results are more preferable than those of our first $\chi^2$ tests using all three transitions. 
Note that the estimated $T_{\rm kin}$ is significantly higher than the $T_{\rm rot}$ estimated via the LTE modeling, 
which indicates that the gas is not thermalized and/or the assumption of optically thin in LTE modeling breaks. 
However, our non-LTE modeling predicts that the high-$J$ lines are optically thin, 
due to the low column density of these states and/or the large velocity widths of these molecules in this galaxy. 

\begin{table*}
  \caption{Parameters used for the $\chi^2$ tests described in the text}
  \label{chisqparams}
  \begin{minipage}{\textwidth}
  \begin{center}
    \begin{tabular}{cccccc}
    \hline\hline
       $R^{\rm{HCN}}_{43/10}$ & $R^{\rm{HCN}}_{32/10}$ & $R^{\rm{HCN}}_{43/32}$ & $R^{\rm{HCO^+}}_{43/10}$ &
       $R^{\rm{HCO^+}}_{32/10}$ & $R^{\rm{HCO^+}}_{43/32}$\\ 
       7.74$\pm$1.10 & 8.36$\pm$1.58 & 0.93$\pm$0.18 & 9.36$\pm$1.36 & 7.97$\pm$1.67 & 1.17$\pm$0.25 \\ 
       & & & & & \\ \hline 
       & & & & & \\
       $R^{\rm{HCN/HCO^+}}_{43/43}$ & $R^{\rm{HCN/HCO^+}}_{32/32}$ & $R^{\rm{HCN/HCO^+}}_{10/10}$ \\
      1.55$\pm$0.43 & 1.96$\pm$0.48 & 1.87$\pm$0.27 & & & \\ \hline
           \end{tabular}
           \footnotetext{{\bf{Notes.}} These values were calculated in [erg cm$^{-2}$ s$^{-1}$], which are used in RADEX. The beam sizes were convolved to that of HCO$^+$ ($J$ = 3--2) (i.e., 4$''$.4 $\times$ 2$''$.7, P.A. = -14.6$^\circ$). Absolute flux error is included.}
    \end{center}
    \end{minipage}
\end{table*}

\section{Molecular gas chemistry in the nucleus of NGC 1097}
In the previous (sub)sections, we presented (1) the high HCN ($J$ = 4--3) to HCO$^+$ ($J$ = 4--3) line ratio ($R_{\rm{HCN/HCO^+}}$ = 2.0$\pm$0.1), 
(2) the high [HCN]/[HCO$^+$] (3.5$\pm$0.6 from the rotation diagram, $\geq$ 5 as preferable values from $\chi^2$ test), 
and (3) that the high-$J$ HCN and HCO$^+$ emission lines seem to be emitted from dense (10$^{4.5}$ cm$^{-3}$ $\leq$ $n_{\rm{H_2}}$ $\leq$ 10$^6$ cm$^{-3}$) 
and hot (70 K $\leq$ $T_{\rm{kin}}$ $\leq$ 550 K) molecular clouds. 
We present some interpretations of these results here. 

\subsection{UV chemistry}
First, we mention that it has been predicted that the fractional abundance of HCN is enhanced through far-UV radiation from hot OB stars (e.g., Tielens \& Hollenbach 1985; Meijerink \& Spaans 2005; Meijerink et al. 2007; Boger \& Sternberg 2005, 2006). 
In the case of NGC 1097, there is some evidence that 
this galaxy has compact young star formation in the nuclear region (e.g., Storchi-Bergmann et al. 2005; Davies et al. 2007).  
However, the star formation rate is estimated to be low ($\sim$ 0.1 $M_\odot$ yr$^{-1}$, calculated from the Table 2 in Davies et al. 2007). 
In addition, the non-detection of PAH features at high angular resolution (Mason et al. 2007) 
and Br$\alpha$ (Kondo et al. 2012) also supports the view of low star-forming activity. 
Apparently, the nuclear star formation in NGC 1097 is much less active than that in nearby starburst galaxies such as NGC 253 and M 82.  
Therefore, it seems difficult to apply the UV chemical models described above, 
which are widely used to explain the chemical properties in starburst galaxies, 
to interpret the HCN-enhancement in NGC 1097. 

\subsection{X-ray chemistry}
Next we suggest that X-ray chemistry can not explain the observations in NGC 1097. 

While FUV photons can only penetrate into the surface region of molecular clouds and create PDRs, 
X-ray photons can deeply penetrate into the central region forming XDRs. 
The difference of the deepness of FUV/X-ray penetration and heating efficiency might thus cause a difference 
of HCN fractional abundance between AGN environments and starburst environments (e.g., Lepp \& Dalgarno 1996; Maloney et al. 1996).
Meijerink \& Spaans (2005) and Meijerink et al. (2007) investigated several molecular abundances 
and predicted intensity ratios of key molecules in both XDRs and PDRs to diagnose these regions. 
While $R_{\rm HCN/HCO^+}$ is $<$ 1 in most of their XDR models, 
it can be $>$ 1 when the effective ionization parameter $\xi_{\rm eff}$ is relatively high, 
i.e., when $F_{\rm X}/n_{\rm H}$ is high and/or $N_{\rm H}$ is low. 
Here, $F_{\rm X}$, $n_{\rm H}$ and $N_{\rm H}$ are X-ray flux, gas density, 
and X-ray attenuating hydrogen column density ($\equiv$ atomic + 2 $\times$ molecular hydrogen column density), respectively. 

We consider these are not applicable to the case of NGC 1097 ($L_{\rm{2-10keV}} = 4.4 \times 10^{40}$ erg s$^{-1}$) because 
(1) $N_{\rm H}$ is at least larger than the molecular hydrogen column density (e.g., $N_{\rm H_2}$ = 4.8 $\times$ 10$^{22}$ cm$^{-2}$ from Section 8.2), 
and would especially be large along the vertical direction to the line of sight considering the torus-like geometry of a type-1 AGN predicted by AGN unified schemes, 
and/or (2) $F_{\rm X}$ is estimated to be very low 
($\sim$ 3.6 erg s$^{-1}$ cm$^{-2}$ at even 10 pc distance from the AGN, assuming the spherical radiation and no attenuation). 
These conditions imply a very low $\xi_{\rm eff}$ in NGC 1097, thus X-ray chemistry can not explain the observations, 
i.e., high $R_{\rm HCN/HCO^+}$, in this galaxy. 

\subsection{High temperature chemistry}
Next, let us reconstruct the chemical layout 
in terms of $``${\it{high temperature chemistry}}$"$, although at this point we ignore the exact nature of the heating source. 
Here, we briefly summarize what kind of chemical reaction is promoted and then what we can expect for the molecular abundance 
in high temperature regions focusing on potential HCN enhancement. 
This kind of chemistry has been proposed in, e.g., Harada et al. (2010, 2013), 
who suggest that the abundance of H$_2$O is enhanced in high temperature environments by the hydrogenation of atomic O as follows:
\begin{equation}
\rm{O + H_2 \longrightarrow OH + H,}
\label{oh}
\end{equation}
\begin{equation}
\rm{OH + H_2 \longrightarrow H_2O + H.}
\label{h2o}
\end{equation} 
This indicates that much of the elemental oxygen is in the form of water, especially at $T$ $\geq$ 300 K, 
and thus the abundances of O and OH are suppressed in such hot regions. 
We consider this water formation to be the trigger in creating the AGN-specific chemical properties. 
The fractional abundances of each molecule in these water-rich environments are discussed below (see details in Harada et al. 2010, 2013). 

\begin{figure*}
  \begin{center}
    \FigureFile(160mm,160mm){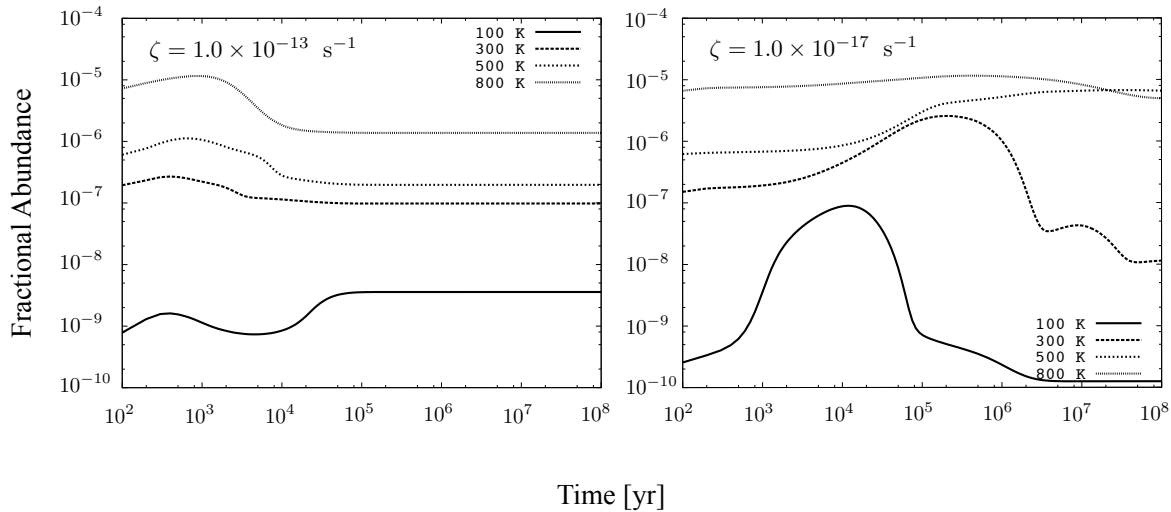}
  \end{center}
  \vspace{-2mm}
  \caption{
  Fractional abundance of HCN as a function of time for a cloud located at the distance of 50 pc from the AGN, 
  with conditions of $n_{\rm H_2}$/$\zeta$ is 3.0 $\times$ 10$^5$/1.0 $\times$ 10$^{-13}$ 
  cm$^{-3}$ $\cdot$ s ({\it{left}}) and 
  3.0 $\times$ 10$^5$/1.0 $\times$ 10$^{-17}$ cm$^{-3}$ $\cdot$ s ({\it{right}}), at various temperatures from 100 K to 800 K. 
  Here $n_{\rm H_2}$ and $\zeta$ are molecular gas density and X-ray ionization rate, respectively. 
  These two panels indicate the expected maximum/minimum $\zeta$ at the distance. 
  The HCN abundance is clearly enhanced as the temperature increases in both panels. 
  See details in Harada et al. (2010).
    		}
		\label{HCNTemp}
 \end{figure*}

{\bf{\underline{HCN:}}} The hydrogenation from CN with H$_2$, 
\begin{equation}
\rm{CN + H_2 \longrightarrow HCN + H,}
\end{equation}
is an endothermic reaction which has a relatively low reaction barrier of $\gamma$ = 820 K, 
thus the HCN abundance is effectively enhanced at high temperatures. 
This description is clearly shown in Figure \ref{HCNTemp}. 
This figure consists of two panels, both presenting the HCN fractional abundance in a molecular cloud 
($n_{\rm H_2}$ = 3.0 $\times$ 10$^5$ cm$^{-3}$; estimated from our RADEX simulation) at 50 pc distance from the AGN in NGC 1097 
($L_{\rm{2-10keV}} = 4.4 \times 10^{40}$ erg s$^{-1}$) as functions of time. 
The left panel is the case of the X-ray ionization rate $\zeta$ = 1.0 $\times$ 10$^{-13}$ s$^{-1}$, which rate is calculated 
by assuming that there is no attenuating material between the AGN and the molecular cloud. 
Therefore, this $\zeta$ is a maximum case at that distance from the AGN. 
On the other hand, the value of $\zeta$ decreases if there exists the attenuating material between the AGN and the molecular cloud, 
and the degree of the decrease depends on the amount of such attenuating material. 
Therefore, we present the case of $\zeta$ = 1.0 $\times$ 10$^{-17}$ s$^{-1}$ in the right panel of Figure \ref{HCNTemp}. 
This value is the canonical cosmic ray ionization rate averaged over the Milky Way (e.g., Spitzer \& Tomasko 1968, Webber 1998), 
and cosmic ray ionization is predicted to affect on the ISM molecular chemistry very similarly to X-ray ionization.  
Therefore, we can use this value (1.0 $\times$ 10$^{-17}$ s$^{-1}$) as a minimum case of $\zeta$ at the molecular cloud we assume here 
because we can expect that the ionization rate is higher in the nuclear region of NGC 1097 than in the average region of the Milky Way. 
Although the actual $\zeta$ in the case of the central 50 pc region of NGC 1097 is not clear 
(but it should be within the maximum and the minimum values above), 
we can expect the fractional abundance of HCN can be enhanced in the case of NGC 1097 at high temperature, 
because such enhancements can be visible in both panels in Figure \ref{HCNTemp}. 
We discuss the nature of the heating source afterwards. 
 
Note that the decrease in O atom due to the formation of water would also lead to 
relatively carbon-rich conditions, which would act to enhance the abundance of other carbonated species such as HCN. 

{\bf{\underline{HCO$^+$:}}} This ion species is generally created as 
\begin{equation}
\rm{CO + H^+_3 \longrightarrow HCO^+ + H_2.}
\end{equation}
However, after the temperature increases and the creation of water and HCN are promoted, 
protonated species of these molecules tend to take up much of the positive charge in the form of H$_3$O$^+$ and HCNH$^+$, 
rather than H$^+_3$ and HCO$^+$, by reactions of, for example, 
\begin{equation}
\rm{H_2O + H^+_3 \longrightarrow H_3O^+ + H_2,}
\label{h3op}
\end{equation}
\begin{equation}
\rm{H_2O + HCO^+ \longrightarrow H_3O^+ + CO,}
\label{hcop_d}
\end{equation}
\begin{equation}
\rm{H_3O^+ + HCN \longrightarrow HCNH^+ + H_2O.}
\label{hcnh}
\end{equation}
Note that H$_3$O$^+$ reforms H$_2$O by dissociative recombinations. 
Reactions such as (\ref{h3op}), (\ref{hcop_d}) and (\ref{hcnh}) would thus prevent the HCO$^+$ creation. 
However, since HCO$^+$ includes both O and C atoms, the effects of less available O would balance with that of more available C, 
which would result in more or less constant relative HCO$^+$ abundance with respect to H$_2$ (though in the severely oxygen-lacked region, it would decrease). 

{\bf{\underline{CS:}}} Here we also mention the fractional abundance of CS, 
because we consider this molecule as a reference frame to estimate the variation of molecular abundances. 
The sulfur chemistry indeed has a close connection to the distribution of reactive oxygen in the gas. 
Detailed chemistry is described in e.g., Charnley (1997), Nomura \& Millar (2004). 
As the water creation is promoted, reactions (\ref{oh}) and (\ref{h2o}) come to dominate the removal of OH and O, 
i.e., most of the reactive oxygen is channeled into water, which is limiting the amount of those species available to take part in the sulfur chemistry. 
This situation leads to an increased abundance of C atoms and destroys SO in the reaction 
\begin{equation}
\rm{SO + C \longrightarrow} CS + O.
\label{cs1}
\end{equation}
This perspective is also supported by the fact that we could not detect any SO and SO$_2$ in our band 3 observation (Kohno et al. in prep.). 
CS is also formed at all temperatures by 
\begin{equation}
\rm{HCS^+} + e^- \rm{\longrightarrow CS + H.}
\label{cs2}
\end{equation}
On the other hand, CS can be destroyed by protonation reactions with HCO$^+$ and/or H$_3$O$^+$, which results in forming HCS$^+$ as 
\begin{equation}
\rm{CS + HX^+ \longrightarrow HCS^+ + X.}
\label{cs3}
\end{equation}
However, there should exist a high amount of electrons in such a highly ionized environment 
as significant amounts of HCO$^+$ or H$_3$O$^+$ do exist. 
This indicates that a significant amount of CS could be reformed by reaction (\ref{cs2}). 
This mechanism would act like a {\it{loop}} between CS and HCS$^+$, 
which would lead to relatively constant abundance of CS with respect to H$_2$ 
in galaxies with different types of activities, as suggested observationally in Mart\'{\i}n et al. (2008, 2009). 
In principle, chemical abundance is determined by the balance of many formation/destruction reactions, 
and some reactions might feed in/out the loop. 
We thus compile more data of species in the chemical chain of sulfur chemistry. 

As described above, once the water creation starts to be promoted due to the high temperature, 
the succeeding water induced reactions will form the HCN enhanced environment 
with respect to HCO$^+$ and possibly to CS, which can explain the observed properties. 
This description is further supported by a detection of enhanced strong H$_2$$S(0)$--$S(3)$/PAH7.7 $\mu$m ratios 
in the nucleus relative to the starburst ring by a factor of two, which indicates the enhanced fraction of 
hot ($T$ $\sim$ 400 K) molecular gas in the nucleus (Beir\~{a}o et al. 2012). 

Then, our next interest is what is the heating source in the nucleus of NGC 1097. 
It is doubtful whether such a low-luminosity AGN in this galaxy 
($L_{\rm{2-10keV}} = 4.4 \times 10^{40}$ erg s$^{-1}$, $L_{\rm{bol}}$ = 8.6 $\times$ 10$^{41}$ erg s$^{-1}$) 
can heat the surrounding ISM to several hundred Kelvin at a tens of parsec scale, 
which is the key to promote the high temperature chemistry and also to produce the H$_2$ emission in the nucleus. 
For example, if we adopt a blackbody approximation, the temperature $T$ [K] at radius $r$ [pc] from the central black hole 
can be calculated by equation (22) in Harada et al. (2010), which results in only 17 K at 40 pc and 106 K at 1 pc. 
Although this is a rather approximate approach, these temperatures are too low to promote high temperature chemistry, 
and thus we conclude that the contribution of X-ray radiation in heating the ISM is not very significant in NGC 1097. 
we need to search for another heating source. 

One other possible mechanism is cosmic ray (CR) heating mainly due to supernovae. 
Beir\~{a}o et al. (2012) estimate a required ionization rate to produce the observed H$_2$$S(0)-S(3)$ emission 
and compare the value with those measured in the Milky Way. 
Then, they conclude it is unlikely that the warm molecular gas in the nucleus of NGC 1097 is heated solely by CR. 
Considering the low star formation rate in the nucleus, CR can not be the dominant heating mechanism in this nucleus. 

Another possible mechanism is mechanical heating. 
Loenen et al. (2008) argued that the PDR, XDR, and even CR chemistry models 
of Meijerink \& Spaans (2005), Meijerink et al. (2006), (2007) could not fully explain the observed line ratios. 
In particular, they pointed out that observed HCN to HNC line ratios ($\equiv$ $R_{\rm HCN/HNC}$) are systematically higher than those predicted by the models. 
The conversion from HNC to HCN is promoted in warm-to-hot gas at temperatures higher than $\sim$ 100 K, 
and they argued such temperatures are only achievable at the edge of molecular clouds in the PDR, XDR and CR models, 
thus we can not expect high $R_{\rm HCN/HNC}$ considering the small volume of such regions. 
To avoid this situation, they added mechanical heating to the models, 
which will turn into the dominant heating mechanism deeper inside the cloud 
because this mechanism is almost independent of depth. 
This approach of combining X-ray and mechanical heating resulted in significantly increased temperature 
and HCN abundance in their model, which could explain the observed line ratios much better. 

Therefore, we searched whether there is some contribution of shock waves in heating the gas in the nucleus of NGC 1097. 
For example, Nemmen et al. (2006) reproduced the nuclear optical to X-ray SED of 
NGC 1097 using a radiatively inefficient accretion flow (RIAF) model. 
Their model indeed underpredicts the radio emission, which suggests the need for another radio continuum source, such as a jet component. 
In addition, a radio to submm SED of the AGN obtained by Matsushita et al. (in prep.) exhibits a typical Gigaherz Peaked Spectrum (GPS), 
which is considered to indicate the existence of a young jet (e.g., Dallacasa et al. 1995; Bickwell et al. 1997). 
Considering these results, mechanical heating due to an AGN jet could be the important heating source. 
However, these is rather circumstantial evidence, and such a jet component or a direct sign of shock heating have not been observed yet. 
Clearly, observations with higher angular resolution and sensitivity are needed to test this hypothesis. 

\subsection{Implications for future observations}
In the previous subsection, we present the high temperature chemistry to explain the HCN enhancement. 
One of the key species of the chemistry is H$_2$O. 
Therefore, we can expect a high abundance of the hydronium ion H$_3$O$^+$ in such warm / hot regions, 
which is formed via H$_2$O (equations (10) and (11)) and also a key species in the chain of oxygen chemistry of dense molecular clouds. 
Its submillimeter rotation-inversion transitions can be used to measure the ionization degree of the gas (Phillips et al. 1992) and trace dense, warm regions selectively. 
Such a kind of study for Sgr B2 in the Galactic center has been shown in van der Tak. et al. (2006). 
Aalto et al. (2011) has demonstrated that the H$_3$O$^+$ to HCO$^+$ column density ratio is much higher in an AGN (NGC 1068) than in starburst galaxies (e.g., NGC 253, IC 342). 
This result is understandable if the positive charge comes in the form of H$_3$O$^+$ rather than as HCO$^+$ due to a high temperature in the AGN. 

Although these submillimeter lines (e.g., $J_{\rm K}$ = 3$^+_2$ -- 2$^-_2$ at 364.7949 GHz, 1$^-_1$ -- 2$^+_1$ at 307.1924 GHz) 
can not be observed efficiently from Mauna Kea due to the atmospheric transmission, they are certainly ideal targets to show the power of ALMA. 
In addition, recent observations with the HIFI heterodyne spectrometer on board {\it{Herschel}} space observatory 
have detected numerous H$_2$O and H$_3$O$^+$ lines around 1 THz (e.g., Wei{\ss} et al. 2010). 
The synergy of ALMA with Herschel will be of great importance to further improve the understanding of the effect of the nuclear activity on the surrounding ISM. 

\section{Summary and conclusion}
In this paper, we present high resolution ($\sim$ 1$''$.5) observations of the 350 GHz band 
including HCN ($J$ = 4--3) and HCO$^+$ ($J$ = 4--3) emission from the nucleus 
of the barred spiral galaxy NGC 1097, which hosts a type 1 Seyfert nucleus, using ALMA band 7. 
This is the first 100 pc scale view of a type-1 AGN obtained by submillimeter dense gas tracers. 
The results of our new observations and conclusions are summarized as follows: 

\begin{itemize}
\setlength{\itemsep}{3mm}
\item[-] The overall structure of the 860 $\mu$m continuum emission is 
	similar to that of the CO ($J$ = 3--2) line emission in the circumnuclear starburst ring. 
	The peak position of the 860 $\mu$m continuum coincides with that of the VLA 6 cm continuum peak. 
	
\item[-] CO emission is distributed in not only the nucleus but also in the circumnuclear starburst ring, 
	which is also observed in CO ($J$ = 2--1) and CO ($J$ = 3--2) (Hsieh et al. 2011). On the other hand, 
	strong HCN ($J$ = 4--3) and HCO$^+$ ($J$ = 4--3) emissions are concentrated primarily towards the nucleus. 

\item[-] We detected CO ($J$ = 3--2), HCN ($J$ = 4--3) and HCO$^+$ ($J$ = 4--3) at $>$ 5 $\sigma$ significance. 
	Other lines including HCN ($v_2$ = 1$^{1f}$, $J$ = 4--3) and CS ($J$ = 7--6), however, were not detected. 
	The HCN ($J$ = 4--3) to HCO$^+$ ($J$ = 4--3) integrated intensity ratio at the 860 $\mu$m peak position is $\sim$ 2.0, 
	which is a high value similar to low-$J$ transitions.  
	
\item[-] Intensity-weighted mean velocity maps of the HCN ($J$ = 4--3) and HCO$^+$ ($J$ = 4--3) line emission 
	show no significant deviation from circular rotation in the nucleus. 
	Assuming that the integrated emission line is broadened by pure circular rotation and a spherical mass distribution, 
	the dynamical mass inside the 40 pc radius is estimated to be 2.8 $\times$ 10$^8$ $M_\odot$. 
	
\item[-] The HCN ($v_2$ = 1$^{1f}$, $J$ = 4--3) to HCN ($J$ = 4--3) integrated intensity ratio in the nucleus is $<$ 0.08, 
	which is at least three times smaller than that observed in NGC 4418 (LIRG). 
	Although we only have a few samples of vibrationally excited HCN so far, especially in extragalactic sources, 
	we infer that IR-pumping is not effective in the nucleus of NGC 1097, 
	and that such a mechanism would not affect the pure rotational population. 

\item[-]  The HCN ($J$ = 4--3) to CS ($J$ = 7--6) integrated intensity ratio, higher than 12.7, was found to be high. 
	A similarly high ratio has also been found in the type-2 AGN, NGC 1068, 
	whereas starburst galaxies do not show such a trend. 
	In addition, in the lower-$J$ case, the HCN ($J$ = 1--0) to CS ($J$ = 2--1) integrated intensity ratio 
	again shows an enhancement in AGNs. 
	Therefore, we suggest the HCN to CS ratio to be higher in AGN-host galaxies than in pure starburst ones, 
	although this is only tentative at this stage. 
	
\item[-] By using ALMA data for $J$ = 4--3 and 1--0, and SMA data for 3--2 transition, we construct HCN and HCO$^+$ 
	excitation diagrams at an angular resolution of 4$''$.4 $\times$ 2$''$.7 (300 pc $\times$ 190 pc). 
	This is the first ever extragalactic high resolution excitation diagram using multiple transitions. 
	The results of a single-component fit are $T_{\rm{rot}}$ = 7.8$\pm$0.3 K, 
	$N_{\rm{HCN}}$ = (2.4$\pm$0.3) $\times$ 10$^{13}$ cm$^{-2}$ for HCN and 
	$T_{\rm{rot}}$ = 8.3$\pm$0.3 K, $N_{\rm{HCO^+}}$ = (6.8$\pm$0.9) $\times$ 10$^{12}$ cm$^{-2}$ for HCO$^+$, respectively. 
	Therefore, the [HCN]/[HCO$^+$] abundance ratio is 3.5$\pm$0.6, which indicates that HCN is more abundant than HCO$^+$. 
	However, our excitation diagrams suggest that there are at least two temperature components in the nucleus. 
	
\item[-] We also ran non-LTE models by RADEX to constrain the excitation parameters ($T_{\rm{kin}}$, $n_{\rm{H_2}}$) of 
	the $J$ = 4--3, 3--2 and 1--0 transitions of HCN and HCO$^+$. 
	We find that only the $J$ = 1--0 transitions show different trends, and conclude that the high-$J$ and low-$J$ emissions 
	are emitted from different regions. 
	The $\chi^2$ tests based on the $R^{\rm{HCN/HCO^+}}_{43}$ and $R^{\rm{HCN/HCO^+}}_{32}$, and the [HCN]/[HCO$^+$] abundance ratio constraints 
	show that the emissions are typically from dense (10$^{4.5}$ cm$^{-3}$ $\leq$ $n_{\rm{H_2}}$ $\leq$ 10$^6$ cm$^{-3}$), 
	hot (70 K $\leq$ $T_{\rm{kin}}$ $\leq$ 550 K) environments. 
	
\item[-] We propose some chemical layouts, and focus especially on high temperature chemistry, to reasonably explain 
	the observed HCN-related properties in this galaxy, although the heating source itself is not clear. 
	
\end{itemize}

We demonstrated that multi-$J$ modeling including HCN and HCO$^+$ by using ALMA observations is a powerful tool 
to investigate the physical/chemical properties of the dense molecular medium around an AGN. 
We will apply this method to other key molecules and other galaxies in future ALMA observations 
in order to promote our research and achieve a comprehensive understanding of ISMs around AGNs. 

\bigskip
{\it{Acknowledgements.}} 
This paper makes use of the following ALMA data: ADS/JAO.ALMA\#2011.0.00108.S (PI: K. Kohno). 
ALMA is a partnership of ESO (representing its member states), 
NSF (USA) and NINS (Japan), together with NRC (Canada) and NSC and ASIAA (Taiwan), 
in cooperation with the Republic of Chile. 
The Joint ALMA Observatory is operated by ESO, AUI/NRAO and NAOJ. 
The National Radio Astronomy Observatory is a facility of 
the National Science Foundation operated under cooperative agreement by Associated Universities, Inc. 
T. Izumi and other authors thank ALMA staff for their kind support. 
K. Fathi acknowledges support from the Swedish Research Council 
and the Swedish Royal Academy of Sciences' Crafoord Prize Foundation. 
S. Mart\'{\i}n acknowledge the cofunding of this work under the Marie Curie Actions of the European Commission (FP7-COFUND). 
A part of this study was supported by MEXT Grant-in-Aid for Specially Promoted Research (No.~20001003). 
We thank the anonymous referee for very helpful suggestions on improving the paper.


\end{document}